\documentclass[review]{elsarticle}
\usepackage{cite} %we use the cite packages
\usepackage{float}
\usepackage{graphicx}
\usepackage{subfig}
\usepackage{graphicx}
\usepackage{amsmath}
\usepackage{booktabs}
\usepackage{lineno,hyperref}
\usepackage{pgfplots}
\usepackage{rotating}
\usepackage{multirow}
\usepackage{lscape}
\usepackage{array}
\modulolinenumbers[5]
\usepackage{tabularx} 
\newcolumntype{C}{>{\centering\arraybackslash}X}
\newcolumntype{R}{>{\raggedleft\arraybackslash}X}
\def \revisionColor {black} %change it to black to turn off
\journal{Digital Investigation}

\makeatletter
\def\ps@pprintTitle{%
 \let\@oddhead\@empty
 \let\@evenhead\@empty
 \def\@oddfoot{}%
 \let\@evenfoot\@oddfoot}
\makeatother
\begin{document}

\begin{frontmatter}

 \title{\textcolor{\revisionColor}{PRNU-Based Source Device Attribution for YouTube Videos}}
 \tnotetext[mytitlenote]{\textcopyright \ 2019. This manuscript version is made available under the CC-BY-NC-ND 4.0 license \ \url{http://creativecommons.org/licenses/by-nc-nd/4.0/}}

%\author{Emmanuel Kiegaing Kouokam}%\author{Elsevier\fnref{myfootnote}}
%\address{kiegaingemmanuel@gmail.com}
%\author{Ahmet Emir Dirik\\}%\author{Elsevier\fnref{myfootnote}}

%\address{edirik@uludag.edu.tr}

\author[firstAuthorAddress]{Emmanuel Kiegaing Kouokam}

\author[secondAuthorAddress]{Ahmet Emir Dirik\corref{mycorrespondingauthor}}
\cortext[mycorrespondingauthor]{Corresponding author}
\ead{edirik@uludag.edu.tr}

\address[firstAuthorAddress]{Department of Electronic Engineering, Uluda\u{g} University, Bursa-Turkey}
\address[secondAuthorAddress]{Department of Computer Engineering, Uluda\u{g} University, Bursa-Turkey}

\begin{abstract}
Photo Response Non-Uniformity (PRNU) is a camera imaging sensor imperfection which has earned a great interest for source device attribution of digital videos. A majority of recent researches about PRNU-based source device attribution for digital videos do not take into consideration the effects of video compression on the PRNU noise in video frames, but rather consider video frames as isolated images of equal importance. As a result, these methods perform poorly on  \textcolor{\revisionColor}{re-compressed or low bit-rate} videos. This paper proposes a novel method for PRNU fingerprint estimation from video frames taking into account the effects of video compression on the PRNU noise in these frames.
\textcolor{\revisionColor}{With this method, we aim to determine whether two videos from unknown sources originate from the same device or not.} 
\textcolor{\revisionColor}{Experimental results on a large set of videos show that the method we propose is more effective than 
existing frame-based methods that use either only I frames or all (I-B-P) frames, especially on YouTube videos}.
\end{abstract}

\begin{keyword}
Video forensics, source device attribution, Photo-Response Non-Uniformity (PRNU), H.264/AVC, YouTube.
\end{keyword}

\end{frontmatter}

%\linenumbers
\section{Introduction}

Digital media (images and videos) are increasingly becoming a popular means for sharing information due to the explosion of smart-phone and tablet sales. Nowadays, people are increasingly using their smart-phones (which they mostly always have at their fingertips) to capture daily life scenes and share them through social media (like Facebook, YouTube, etc.). Apart from being a formidable means to communicate or share emotions, digital media can also be used to perpetrate crimes such as movie piracy, terrorist propaganda or child pornography. Furthermore, digital images or videos can be used as legal evidence during a trial in a court of justice. For these reasons, multimedia forensics is increasingly attracting the attention of forensic scientists and government agencies.

\hspace{0.12cm} Identifying the device from which a digital media originates can sometimes be very crucial to an investigation. For instance, it can mean that the owner of the camera witnessed the scene that was captured and that he was at the place where the footage was taken. Determining the source device of an image or video during a trial in a court of justice or digital investigation can help to incriminate a suspect (for instance, a pedophile) possessing or sharing photos/videos acquired with the same camera device. \textcolor{\revisionColor}{Sometimes the camera device is not physically available. In this type of situation, query/input images or videos can be matched with another set of images or videos \textcolor{\revisionColor}{seized} during the investigation to check whether their source cameras are the same or not.} Such an analysis can be done through meta-data (EXIF) of the subjected images or videos. However, sometimes EXIF data are removed by third-party applications either intentionally or while sharing their contents on a social network.
A different (yet more effective) source camera attribution method consists of examining specific \textcolor{\revisionColor}{noise patterns that are present in acquired images or videos due to sensor imperfections}. The idea of using CCD (Charge-Coupled Device) sensors' imperfections to perform video source identification first originated from K. Kurosawa late in 1999. In \citep{Kurusawa1999}, he showed that dark currents in the CCD chips of camcorders form a fixed noise pattern which is added to recorded videotapes. This fixed noise pattern is used as a "fingerprint" to identify the source device of a given recording.

Jan Lukas et al. in \citep{Jan2005} showed that, as an intrinsic, natural, and unique camera fingerprint, the Photo-Response Non-Uniformity (PRNU) noise in digital images could effectively be used to perform digital image source attribution and forgery detection. This seminal work on PRNU-based image source attribution was followed by many others which established the PRNU as one of the most promising and powerful imaging sensor \textcolor{\revisionColor}{characteristics} which can be exploited for image source attribution.

Chen et al. in \citep{MoChen2007} investigated the video source device attribution problem and showed that PRNU could effectively be used to identify the source camcorder of a subjected digital video (even for low-resolution cases) by estimating the PRNU fingerprint or sensor pattern noise (SPN) from individual video frames given that enough frames are available (a video clip of ten minutes was sufficient to identify the source device of low-resolution videos such as 264$\times$352 pixels). Dai-Kyung et al. in \citep{Dai2012} improved the results in \citep{MoChen2007} by applying a MACE (Minimum Average Correlation Energy) filter to the reference PRNU fingerprint while testing its similarity with a query video's sensor pattern noise (SPN). Through this, an improvement of up to 10\% of the decision accuracy was achieved compared to Chen's method for relatively small video resolutions such as 128$\times$128 pixels.

\textcolor{\revisionColor}{W. van Houten and Z. Geradts} in \citep{Wiger2009} investigated the usage of PRNU for source attribution of YouTube videos. A set of webcams and codecs were used to record and encode videos. These videos were later uploaded to and downloaded from YouTube. SPN was then estimated from downloaded videos and used for source device attribution. Even though this work gave good results, 
\textcolor{\revisionColor}{ its findings, which date to 2009, are out of date because the video cameras (notably handheld devices) used by YouTube users have considerably evolved since then.}

Louis Javier et al. proposed in \citep{Luis2016} a video source identification scheme based on the usage of PRNU and Support Vector Machines (SVM). A set of 5 smart-phones from 5 different brands were used to acquire videos used in training and testing steps. A total of 81 features, which are the SPN wavelet components, were used to feed the SVM classifier. Only native videos (videos taken from the acquisition devices without any post-processing) which have been cropped to various resolutions were used in the experiments. It was reported that the proposed classification scheme had an accuracy of about 87\% to 90\% depending on the video resolution. In general, features-based classification algorithms are not suitable for source attribution problems because there could be millions of devices that have the same brand and model. In such cases, a forensic examiner should model each of these devices as a separate class which is not practical in real life scenarios.

Massimo et al. in \citep{Massimo2017} proposed a ``hybrid" approach to video source attribution. Retaking the idea of utilizing still images for camera fingerprint estimation that was previously introduced in \citep{Samet2016}, in \citep{Massimo2017}, they established camera specific transfer functions between fingerprints estimated from images and video frames from the same camera for a broad set of smart-phone and tablet cameras (this is called image-to-video matching). These transfer functions consist of crop and scale parameters that best match these two fingerprints that have different resolutions and aspect ratios. The camera-specific transfer function is applied to the fingerprint estimated from still images before correlating it with the SPN estimated from the frames of a (non-stabilized) query video for source device attribution. This approach also solves the problem of estimating PRNU fingerprint of cameras featuring digital video stabilization (like iPhones and some Android smartphones) since this camera feature misaligns the PRNU noise from one frame to another as it has been stated in \citep{Samet2016}. Massimo et al. also \textcolor{\revisionColor}{discussed how} to link a Facebook account to a YouTube account by correlating two PRNU fingerprint estimates obtained from a query video downloaded from YouTube and images shared on a specific Facebook account, but the accuracy of by their method was very low.

The method in \citep{Massimo2017} has good identification results on native videos but source attribution accuracy for YouTube videos are not as high as for native camera outputs. Moreover, this method cannot be used to perform video-to-video device linking for cameras featuring video stabilization. Furthermore, in the case of Facebook-shared images, estimating a fingerprint using images from an unknown source is not realistic since it is assumed that they all come from the same device, which may not always be the case.

In video source device attribution, it is very crucial to estimate the PRNU fingerprint accurately.  The original version (native camera output) of a query video is not always accessible during an investigation. In some cases, only a resized, re-compressed, or cropped version of the video is available for forensic examination. In most of the previous researches related to PRNU-based source device attribution for digital videos, the effects of video compression are not taken into account when estimating the PRNU fingerprint from video frames. Some authors like Samet et al. in \citep{Samet2016} just assume that I frames are the best to be used, others like Dasara et al. in \citep{Dasara2017} give equal importance to I, P and B frames, and use all video frames for fingerprint estimation. Accordingly, they reported that a low accuracy in source attribution is obtained when performed on videos re-compressed by YouTube or Whatsapp (compared to their native version). Thus, it is obvious that video compression significantly affects or degrades the PRNU noise in video frames. This fact has to be taken into account when estimating the PRNU fingerprint from highly compressed videos.

In this study, we show the limits of the above-mentioned approaches (utilizing I or all frames) for fingerprint estimation by testing them under different scenarios (Table \ref{table:experience cases}) for native and YouTube video cases. We will call these approaches ``frame-based" in the rest of the paper.

In the paper, we will briefly describe the H.264/AVC video compression standard, then study the operations applied on an encoded frame block and investigate how PRNU noise is affected locally by these operations. Accordingly, we will propose a novel method for PRNU fingerprint estimation which takes into account the effects of video compression on PRNU noise in video frames. We call this method ``block-based" approach since it relies on block-wise in-frame noise analysis. The proposed and frame-based methods are tested with a wide range of videos available in the VISION database \citep{Dasara2017} acquired from various smart-phones and tablets (Table \ref{tab:listdevices}). 

We will particularly test a scenario where two query videos (native or YouTube) are compared with each other based on their estimated SPN to determine whether they originate from the same source or not. It should be noted that neither EXIF nor any side information except the estimated SPN of videos is used in the analysis.

The rest of the paper is organized as follows: Section 2 introduces PRNU-based image source camera attribution for still images. Section 3 presents the principles of H.264/AVC video compression and its impact on the PRNU noise estimation from video frames. Section 4 introduces the frame-based and the block-based approaches  for source video device attribution in detail. Section 5 provides the details of the experimental setup. The experimental results are presented in Section 6. Finally, Section 7 concludes the paper and discusses future works.

\section{PRNU-based Source Camera Attribution} 
The camera sensor is at the heart of the image acquisition process. It is made of a large number of small photo-detectors called pixels. Pixels use the photoelectric effect to convert incident light (photons) to electrons. For a given intensity of light falling on a pixel, the amount of electrons generated depends on the pixel's physical dimensions and silicon homogeneity. Because of the imperfections of the manufacturing process and the non-homogeneity naturally present in the silicon, all the pixels of a sensor will never have the same photo-response characteristics. This phenomenon is called Photo-Response Non-Uniformity, and it is inevitable for all type of camera sensors (both CCD and CMOS). 

Let the PRNU of an imaging sensor be represented by a matrix \textbf{K}, having the same dimensions with the sensor. A simplified and linearized imaging sensor model \citep{Jessica2009} can be written as: 
\begin{equation}
\mathbf{I} = \mathbf{I}^{(0)} + \mathbf{I}^{(0)}\mathbf{K} + \mathbf{\Psi}
\label{eqn: sensor model}
\end{equation}
where \textbf{I} represents the sensor output, $\mathbf{I}^{(0)}$ is the ideal sensor output in the absence of any noise, $\mathbf{I}^{(0)}\mathbf{K}$ is the sensor's PRNU, and $\mathbf{\Psi}$ is the temporal random noise  comprising of thermal noise, shot noise, and other noise components. The matrices in (\ref{eqn: sensor model}) have the same size. Throughout the paper, all the mentioned matrix operations are element-wise. 
The PRNU noise is non-temporal, random, and unique to \textcolor{\revisionColor}{each} camera sensor. It is pretty robust to lossy JPEG compression \citep{MoChen2008}  and  \textcolor{\revisionColor}{global image scaling \citep{Miroslav2008}}. These properties make PRNU a reliable quantity (intrinsic camera fingerprint) which can be used to perform many digital image forensic tasks such as source device identification, device linking, and forgery detection \citep{Jessica2009}.

\subsection{PRNU fingerprint estimation}
The PRNU noise pattern of an imaging sensor can be estimated through a set of images of the same camera device. Having a number $d$ of images of the same camera, the camera PRNU fingerprint is estimated with a maximum likelihood estimator \citep{Jessica2009} as follows:
\begin{equation}
\mathbf{F} = \frac{\sum\limits_{k=1}^d \mathbf{W}_{k}\mathbf{I}_{k}}{\sum\limits_{k=1}^{d}(\mathbf{I}_{k})^{2}}
\label{eqn:K estimate}
\end{equation}
where $\mathbf{I}_{k}$ is the $k$ th image acquired from the same camera device. 
$\mathbf{W}_{k} = \mathbf{I}_{k} - \textrm{Denoise}(\mathbf{I}_{k})$ is the difference between the original image $\mathbf{I}_{k}$ and its denoised version. The denoised version of the image $\mathbf{I}_{k}$ is obtained using a wavelet-based denoising filter as described in \citep{Mihcak1999}. For color images, three fingerprints corresponding to the three color channels (red, green, and blue) are estimated separately then combined like in a generic RGB to gray conversion \citep{Jessica2009}.

The estimated fingerprint $\mathbf{{F}}$ is made of two components: the reference pattern (RP) and the linear pattern (LP). The linear pattern contains all the noise components that are systematically present in an image due to artifacts introduced by Color Filter Array (CFA) interpolation, JPEG compression, and post-processing operations performed in the image acquisition pipeline. Contrary to the reference pattern, the linear pattern is common to the cameras of the same model; thus, it has to be removed from the fingerprint to achieve an accurate source attribution even with cameras of the same model. The linear pattern can be used to identify a camera model as it was done in \citep{Thomas2008}. Removing the linear pattern from the fingerprint is a straightforward task since it appears periodically in $\mathbf{{F}}$. In \citep{Jessica2009}, the linear pattern is removed from the fingerprint by subtracting the averages of each row and column from the corresponding element in $\mathbf{{F}}$. The estimated fingerprint is then filtered with a Wiener filter in the DFT domain to meet the zero mean Gaussian white noise model hypothesis.

It has been stated in \citep{Jessica2009} that 20 to 50 natural (any content) images are enough to obtain a good estimate of a camera's fingerprint. However, a fingerprint with similar accuracy can also be obtained \textcolor{\revisionColor}{using 10 to 25 images with flat content (blue skies or flat walls).}

\subsection{Source camera attribution} 
\label{section : image source identification}
Source camera attribution of a query image having an estimated PRNU noise $\mathbf{W}$ and a camera having a PRNU fingerprint \textbf{F} is formulated as a two-channel hypothesis testing problem as follows \citep{Miroslav2008}:
\begin{equation}
\begin{array}{lcl} H_{0}:\mathbf{F} \neq \mathbf{W}  \\ H_{1}:\mathbf{F} = \mathbf{W} \end{array}
\label{eq:test Hypothesis}
\end{equation}
This hypothesis can be tested by taking normalized cross-correlation of the noise and the fingerprint estimates as:
%NCC equation
\begin{equation}
\displaystyle \mathbf{\rho}(r,c) =  \frac{\sum\limits_{i=1}^{m}\sum\limits_{j=1}^{n}(\mathbf{F}(i,j) - \overline{\mathbf{F}})(\mathbf{W}(i+r-1,j+c-1) - \overline{\mathbf{W}})}{||\mathbf{F} - \overline{\mathbf{F}}||||\mathbf{W} - \overline{\mathbf{W}}||}
\label{eq:Ncc}
\end{equation}
where $\mathbf{\overline{F}}$ and $\mathbf{\overline{W}}$ represent the averages of $\mathbf{F}$ and $\mathbf{W}$, respectively. The operator $|| \text{ } ||$ is the Euclidean norm, \textit{r} and \textit{c} are circular shift parameters ranging from 1 to \textit{m} and 1 to \textit{n}, respectively. We assume that $\mathbf{F}$ and $\mathbf{W}$ have the same size $m \times n$. The Peak to Correlation Energy (PCE), a resolution independent similarity metric, is computed from normalized cross correlation (NCC) as follows:
\begin{equation}
\mathrm{PCE}(\mathbf{\rho})= \frac{\rho_{peak}^{2}}{\frac{1}{\mathit{mn}-|S|}\sum\limits_{r,c \not\in S}{\rho(r,c)^{2}}}
\label{eq:PCE equation}
\end{equation} 
where $\rho_{peak}$ is the maximum value of NCC matrix, $S$ is a small region surrounding $\rho_{peak}$ and $|S|$ is the cardinality of $S$. When matrix resolutions of the fingerprint and the noise estimate are the same, $\rho_{peak}$ can be replaced directly with $\rho(1,1)$. If $\mathrm{PCE}(\mathbf{\rho})$ is above a decision threshold $\mathbf{\tau}$, the null hypothesis ($H_{0}$) is rejected and the query image with the noise estimate $\mathbf{W}$ is assumed to be acquired with the same camera of the fingerprint $\mathbf{F}$.

\section{The H.264/AVC Video Compression}\label{section: video compression basics}
This section presents key aspects of the H.264/AVC (Advanced Video Compression) video compression standard and shows how operations involved in video compression affect the PRNU noise in frame blocks. The H.264/AVC video compression standard is the world's leading standard for video compression. Nowadays, it is used by almost all smart-phones and video-sharing platforms (or social media) like YouTube and Facebook. The H.264/AVC standard is managed by the JVT (Joint Video Team). Its first version was released in 2003 and is destined to be replaced by the H.265/HEVC (High-Efficiency Video Coding) standard in the next decade. An exhaustive description of video coding techniques is out of the scope of this paper, and the reader can refer to \citep{Richardson2010} for a comprehensive description of the H.264/AVC video coding standard and \citep{IUT2009} for technical details.

\begin{figure}
\begin{center}
\includegraphics[scale=0.48]{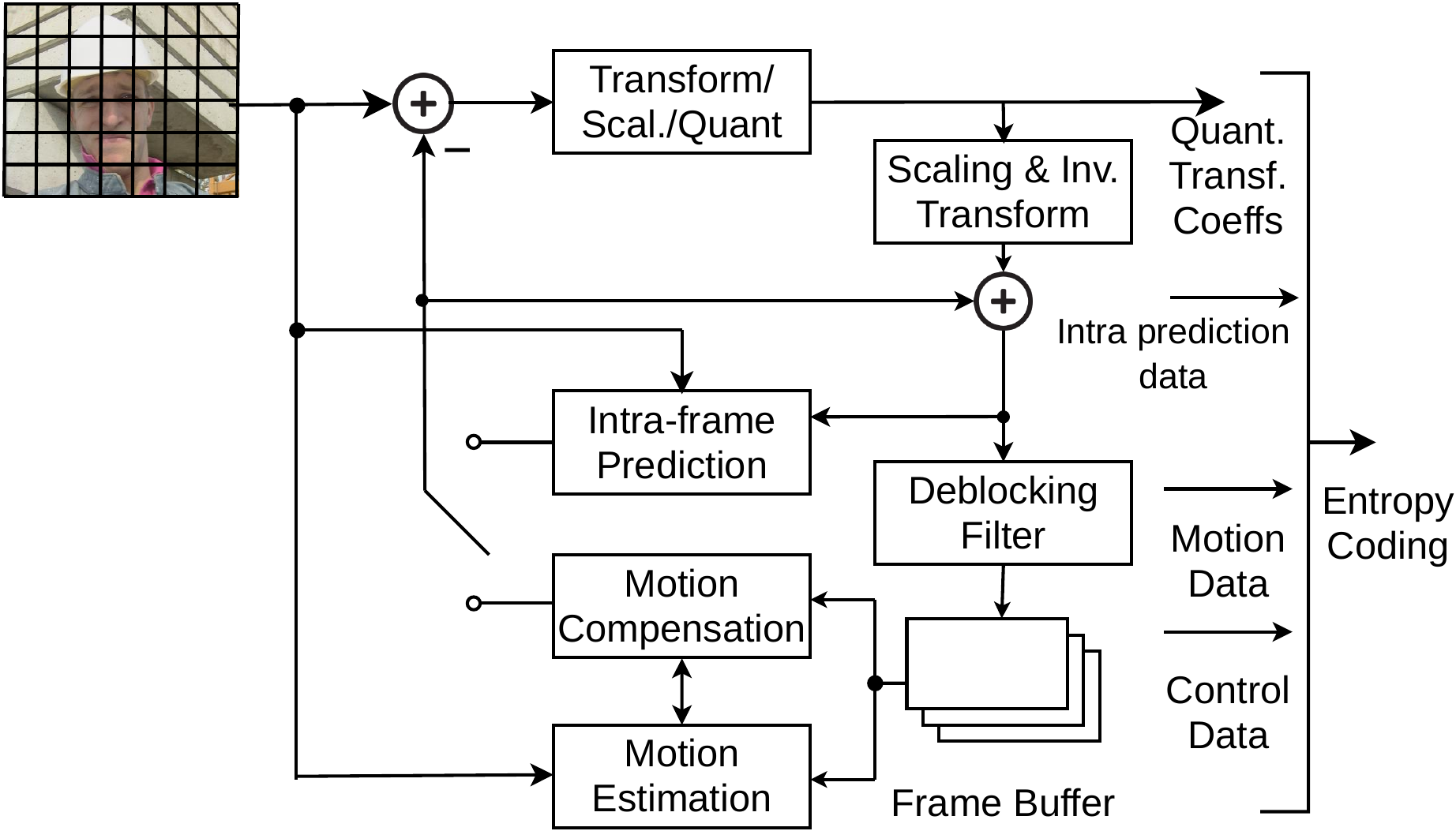}
\end{center}
\caption{Simplified diagram of a H.264/AVC encoder}
\label{fig:h264 Encoder}
\end{figure}

A simplified diagram of an H.264/AVC encoder is given in Fig. \ref{fig:h264 Encoder} \citep{Richardson2010}. A video encoder also embeds a decoder. Modern video compression standards share several key operations such as block processing, prediction, transform, quantization, entropy coding.
\begin{itemize}
\item \textit{Block processing}: The input frame is divided into one or more slices containing Macro-blocks of size 16$\times$16. These Macro-blocks are divided into blocks of different sizes (16$\times$16, 8$\times$16, 8$\times$8, 4$\times$8, 4$\times$4 ...) according to the type of prediction used to encode them. Subsequent operations such as prediction, transform, and quantization are performed on these sub-blocks.
\item \textit{Prediction}: Prediction is a process in which a current block's pixels are predicted from pixels of a previously encoded block(s) within the current frame (intra-frame coding) and/or previous or future encoded frames (inter-frame coding).
After prediction, the prediction residue (the difference between the current block and the predicted block) is computed.
\item \textit{Transform}: A transform operation is applied on block prediction residue and aims at reducing the statistical correlation between its samples such that most of the information it contains can be concentrated into a small number of encoded samples. The H.264/AVC standard uses (integer) Discrete Cosine Transform (DCT) with integer transform cores of size 4x4 or 8x8 (used exclusively in High profile encoders).
\item \textit{Quantization}: Quantization consists of reducing the precision used to represent sample values. It aims at reducing the number of bits necessary to represent a set of values. In H.264/AVC, each Macro-block has its quantization parameter which can be a scalar or a quantization matrix like in JPEG (used only in High profile encoder). It is important to note that among all the operations involved in video compression, quantization is the only operation which is non-reversible. 
\item \textit{Entropy coding}: Entropy coding is a process through which discrete-valued symbols are represented in a manner that takes advantage of the relative probability of each source symbol. In H.264/AVC, VLC (Variable length coding) or arithmetic coding (CABAC) can be used for entropy coding.
\end{itemize}

Prediction has brought the greatest increase in coding efficiency to the H.264/AVC compression standard in comparison to the previous coding standards (like MJPEG). Prediction in video compression exploits the spatial and temporal redundancies highly present in video sequences. Spatial redundancy is exploited in intra-frame prediction. In intra-frame prediction, prediction (reference) blocks and blocks to be predicted are all located in the same frame (neighboring blocks). Intra-frame predicted Macro-blocks are called I Macro-blocks. Temporal redundancy is exploited in inter-frame prediction and is based on motion estimation and motion compensation. In inter-frame prediction, a Macro-block is predicted using Macro-blocks in past and future frames. There are mainly two types of inter-frame predicted Macro-blocks in H.264/AVC: P Macro-blocks (which are predicted using only blocks in past frames) and B Macro-blocks (which are predicted using blocks in past and the future frames). A de-blocking filter is applied to inter-frame predicted blocks to reduce block artifacts due to motion compensation.
 In an H.264/AVC encoded video, three types of frames can be found: I, P, and B frames. An I frame is made only of I Macro-blocks, a P frame of I and P Macro-blocks and, a B frame of I, P, and B Macro-blocks. 
 
\subsection{The impact of video compression on PRNU noise}

As we see notice, video compression is by far more complex than still image compression. Thus, the statement made in \citep{MoChen2008} according to which the PRNU noise survives lossy JPEG compression might not hold for H.264/AVC compression given that operations applied to frame blocks during compression also affect the PRNU noise they contain. Here, we determine the condition necessary for the PRNU noise in an encoded block to survive video compression. 

\begin{figure}
\begin{center}
\includegraphics[scale=0.52]{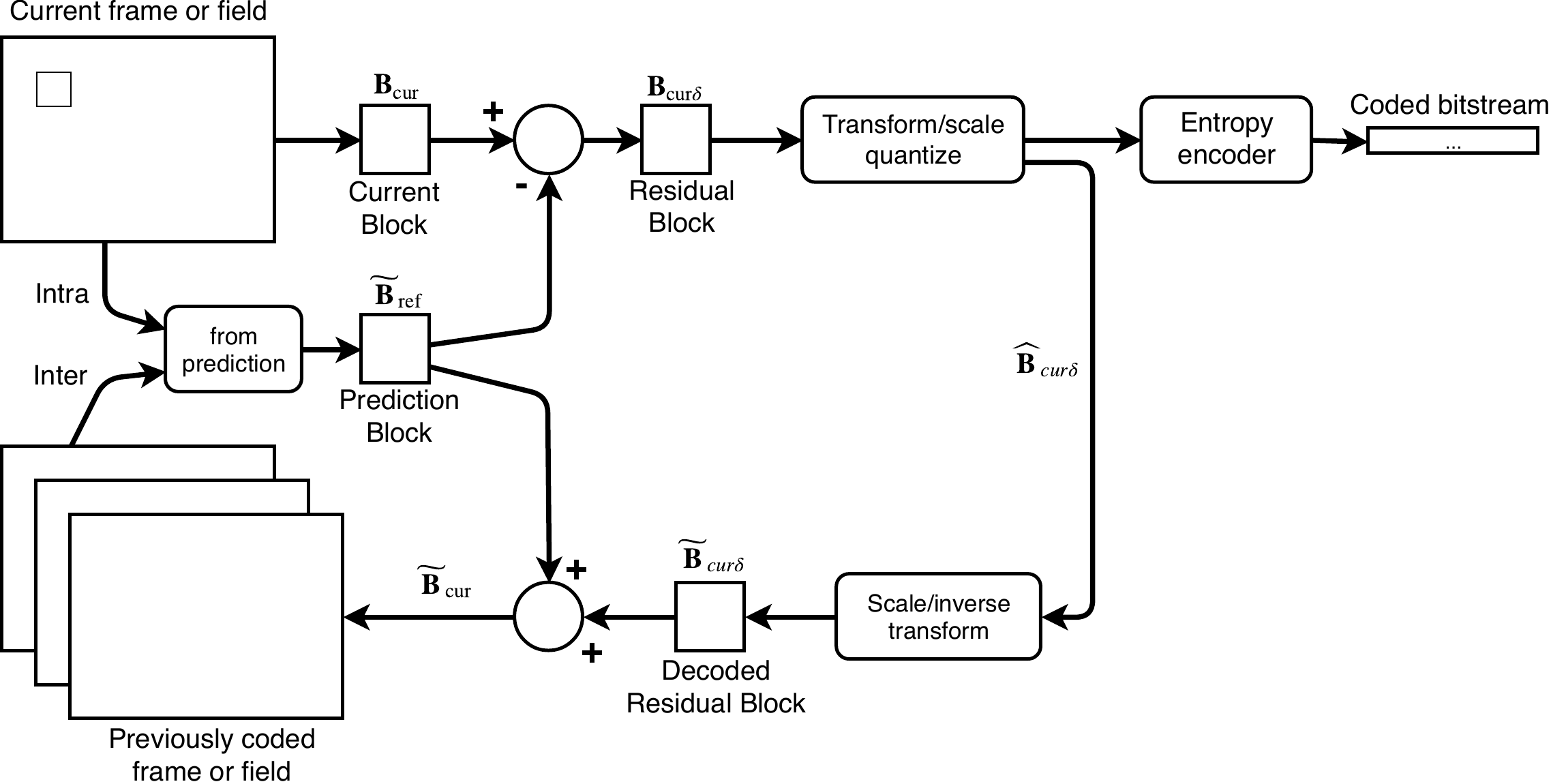}
\end{center}
\caption{Operations applied on a frame block during encoding/decoding}
\label{fig:operations on block prnu}
\end{figure}

%As previously mentioned, the only non-reversible operation involved in video compression is (residue) quantization. 
To investigate the effects of video compression on the PRNU noise in encoded frame blocks, let us consider a block which is to be encoded. We note: $\mathbf{B}_{cur}$ as the block which is to be encoded, $\mathbf{\widetilde{B}}_{cur}$  as the decoded current block, $\mathbf{\widetilde{B}}_{ref}$ as the reference (or prediction) block (a previously encoded and decoded block), $\mathbf{B}_{cur\delta}$ as the current block's prediction residue (see Fig. \ref{fig:operations on block prnu}). The operations applied to $\mathbf{B}_{cur}$ during its encoding and decoding processes are given by equations (6) to (8). 

The prediction residue is computed as: 
\begin{flalign}
\mathbf{B}_{cur\delta} = \mathbf{B}_{cur} - \mathbf{\widetilde{B}}_{ref}&
\end{flalign}
The output of the transform (DCT), scaling, and quantization operations for  the residue input can be written as:
\begin{flalign}
\mathbf{\widehat{B}}_{cur\delta} = \mathrm{Quant}[\mathrm{Scale}[\mathrm{DCT}(\mathbf{B}_{cur\delta})]]&
\end{flalign}
Finally, the block's decoding equation can be written as:
\begin{flalign}
\mathbf{\widetilde{B}}_{cur} = \mathbf{\widetilde{B}}_{ref} + \mathbf{\widetilde{B}}_{cur\delta} = \mathbf{\widetilde{B}}_{ref} + \mathrm{DCT}^{-1}[\mathrm{Scale}(\mathbf{\widehat{B}}_{cur\delta})]
\label{eq: block reconstruction equation}
\end{flalign}

Equation \ref{eq: block reconstruction equation} shows that the content of a decoded block $\mathbf{\widetilde{B}}_{cur}$ is highly dependent on the inverse DCT transform of the block's prediction residue. A 2D-DCT transform matrix is composed of two classes of coefficients: a DC coefficient which is the one at the location (0,0) in the matrix, and, AC coefficients which are the remaining. That being said, it is easily noticed that the high-frequency (AC) content of an encoded block is lost if its	prediction	residue's DCT-AC coefficients are	all	zero. In such a case, the high-frequency content of the decoded block is (approximately) equal to the one in its reference block(s). Based on this, we can conclude that the PRNU noise in an encoded block is not (completely) destroyed by video compression if the DCT-AC coefficients of its prediction residue are not all null (because the PRNU noise is by essence a high frequency signal). The strength of the PRNU noise remaining in a decoded block depends on the number of its non-null DCT-AC coefficients and on the scene content. It is shown in \citep{Chang2010} that high-frequency content scenes (notably edges) interfere with the PRNU noise estimation. If the DCT-AC coefficients of a given block's prediction residue are all null, then the PRNU noise it contains is irreversibly lost and replaced by the one in its prediction block(s).

\section{Proposed Method: Compression-Aware PRNU Estimation }
\label{sec:proposed}
In this section, we first present the strategies which are currently used for camera fingerprint and video noise estimation from video frames.  Secondly, we introduce a new approach called the block-based approach for highly compressed videos taking into account the effects of video compression that have been discussed in the previous section.

\subsection{Frame-based approach for video source attribution}
As it is mentioned in Section \ref{section: video compression basics}, an H.264/AVC compressed video is made of three types of frames: I, P, and B frames. For video source attribution, it is suggested in the literature to use only I frames assuming that they are significantly less compressed than B and P frames because intra-frame coding is solely used in I frames. Thus, intuitively, I frames seem to be the best set of frames to use for PRNU fingerprint estimation. But, is this practice better than using all video frames (I, B, and P frames) for PRNU fingerprint estimation with regard to source attribution accuracy? To answer this question, we test the two cases presented in Table \ref{table:experience cases} where only I frames are used (case-1) and where all the frames in videos are used (case-2). In these two cases, a given camera's PRNU fingerprint $\mathbf{F}_{v}$  (estimated from a flat-content video) and a given natural-content video's SPN $\mathbf{W}_{v}$ are all computed using the Equation (\ref{eqn:K estimate}). We also consider the case where we do not have the suspect camera device, but we want to figure out whether two query videos have the same source or not (device linking). In this case, we cannot have a reference fingerprint and thus we simply match the SPN matrices estimated from each query video using (\ref{eqn:K estimate}).

\begin{table}
\begin{center}
\begin{tabular}{ccc}
%\toprule
\hline 
\hline 
Case & Fingerprint estimated from & Video noise estimated from\\
\hline 
$\textrm{C}_1$ & I frames & I frames\\
%\hline 
$\textrm{C}_2$ & I + P + B frames  & I + P + B frames \\
\hline 
\hline  
\end{tabular} 
\end{center}
\caption{Different cases of fingerprint and noise estimation in the frame-based approach}
\label{table:experience cases}
\end{table} 

\subsection{Block-based approach for video source attribution}
The block-based approach goes further than just selecting the frames to be used for fingerprint or noise estimation. In each frame, we seek particular blocks in which the PRNU noise has not been completely degraded by video compression. As we have \textcolor{\revisionColor}{discussed} earlier, the PRNU noise in a block survives compression if the block's prediction residue DCT-AC coefficients are not all zero. Thus, to estimate the video PRNU noise, we will only use blocks which have at least one non-null DCT-AC coefficient in I, P, and B frames since they still have (at least partially) some amount of PRNU noise at its correct location/block. %\textcolor{\revisionColor}{Setting the block selection threshold to one (number of non-null DCT-ACT coefficients $\geq$ 1) is the best option since a bigger threshold value would discard blocks that have a small amount of PRNU noise which could improve the overall PRNU noise estimation; see  (\ref{eq:Mk}).} 

The block-based approach requires to analyze all the DCT-AC coefficients of all the frame-blocks of an \textcolor{\revisionColor}{input} video to determine the appropriate blocks for PRNU noise estimation. For each frame of an \textcolor{\revisionColor}{input} video, the DCT-AC coefficients of blocks' prediction residue are read and checked whether they are all zero or not. This task is carried out by a modified version of the H.264/AVC reference decoder jm16.1 \citep{jm2015}. We assign labels (1 or 0) to all frame-pixels to indicate that the frame-block they are positioned in has some PRNU noise at its correct location (label:1) or not (label:0). As a result, we obtain a binary matrix (frame mask $\mathbf{M}$) for each frame which indicates the appropriate pixels/blocks to be used in the frame-wise PRNU noise estimation. The zeros in the frame mask indicate the location of pixels/blocks where PRNU noise estimation is not feasible. An element of the frame mask $\mathbf{M}$ of any $k$ th frame at pixel location $(r,c)$ is computed according to (\ref{eq:Mk}). The frame masks $\mathbf{M}_k$ are of the same dimensions with the investigated video's resolution. As we have mentioned earlier, the H.264/AVC standard uses 4$\times$4 or 8$\times$8 (in High profile encoder) sized integer-DCT core transforms. A flag in Macro-blocks' header indicates the size of the core used.  
\begin{flalign}
  M_k(r,c)= 
  \begin{cases} 
  0 {\textrm{ , if DCT-AC coefficients at $(r,c)$ are all \textit{zero}}} \\
  1  {\textrm{ , \textcolor{\revisionColor}{otherwise}}}
  \end{cases}
  \label{eq:Mk}
\end{flalign}

\textcolor{\revisionColor}{
If there is at least \textit{one} non-zero DCT-AC coefficient in a particular frame-block, we set $M_k(r,c)= 1 $ and use the PRNU noise of that block in video noise estimation. Otherwise, we set $M_k(r,c)=0$ which discards the block during video noise estimation. If we set the number of non-zero DCT-AC coefficients threshold to  values higher than  \textit{one} (2,3,...), we would discard some frame-blocks that have some amount of proper PRNU noise during video noise estimation. As a result, the accuracy of the source device attribution could decrease.}

\begin{figure}
%\begin{table}[]
\setlength\tabcolsep{0.5pt}
%\setlength\tabrowsep{0pt}
%trim=left bottom right top
    \centering
    \begin{tabular}{ccccc}
      \includegraphics[scale=0.15]{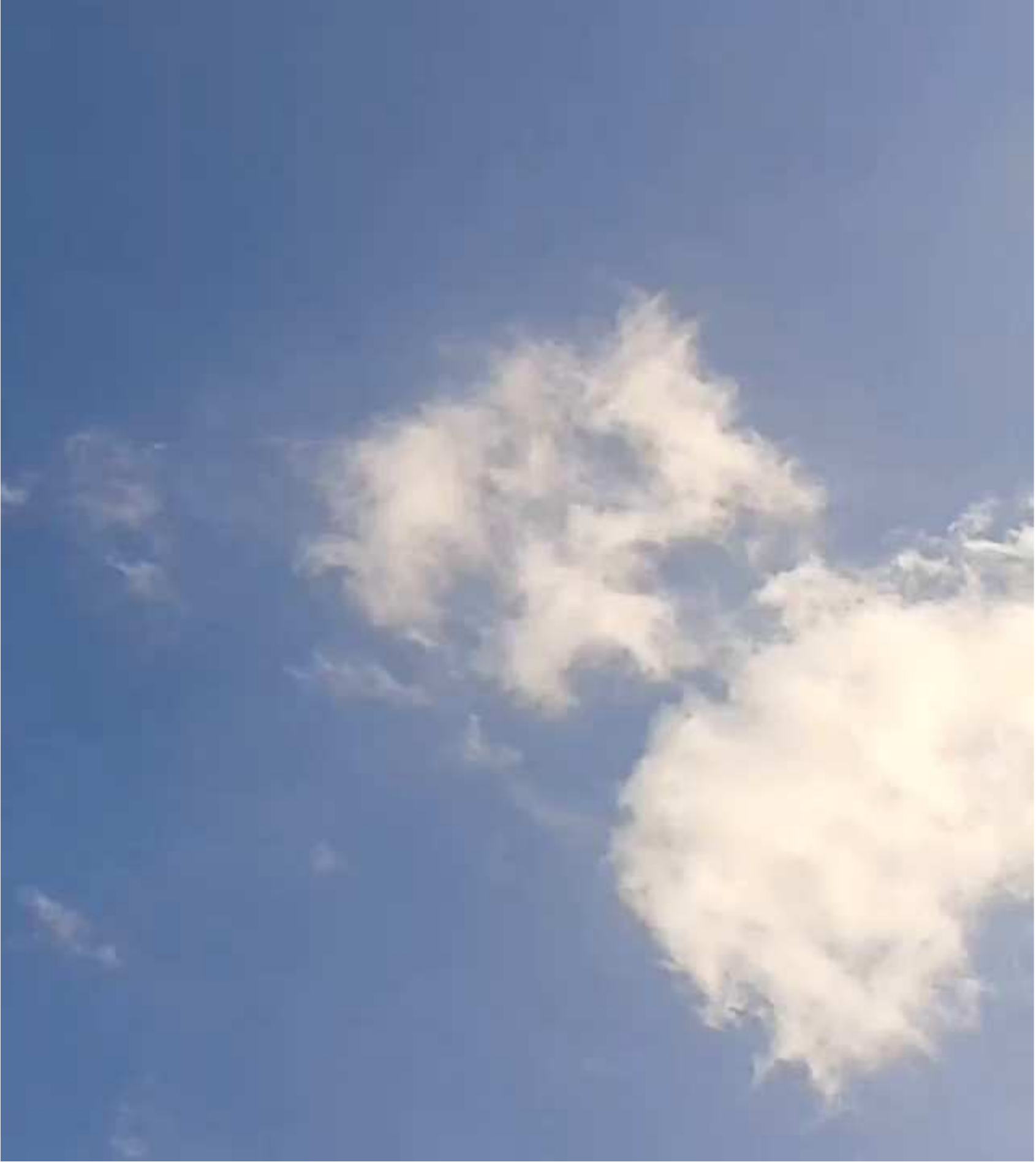}  & 
      \includegraphics[scale=0.15]{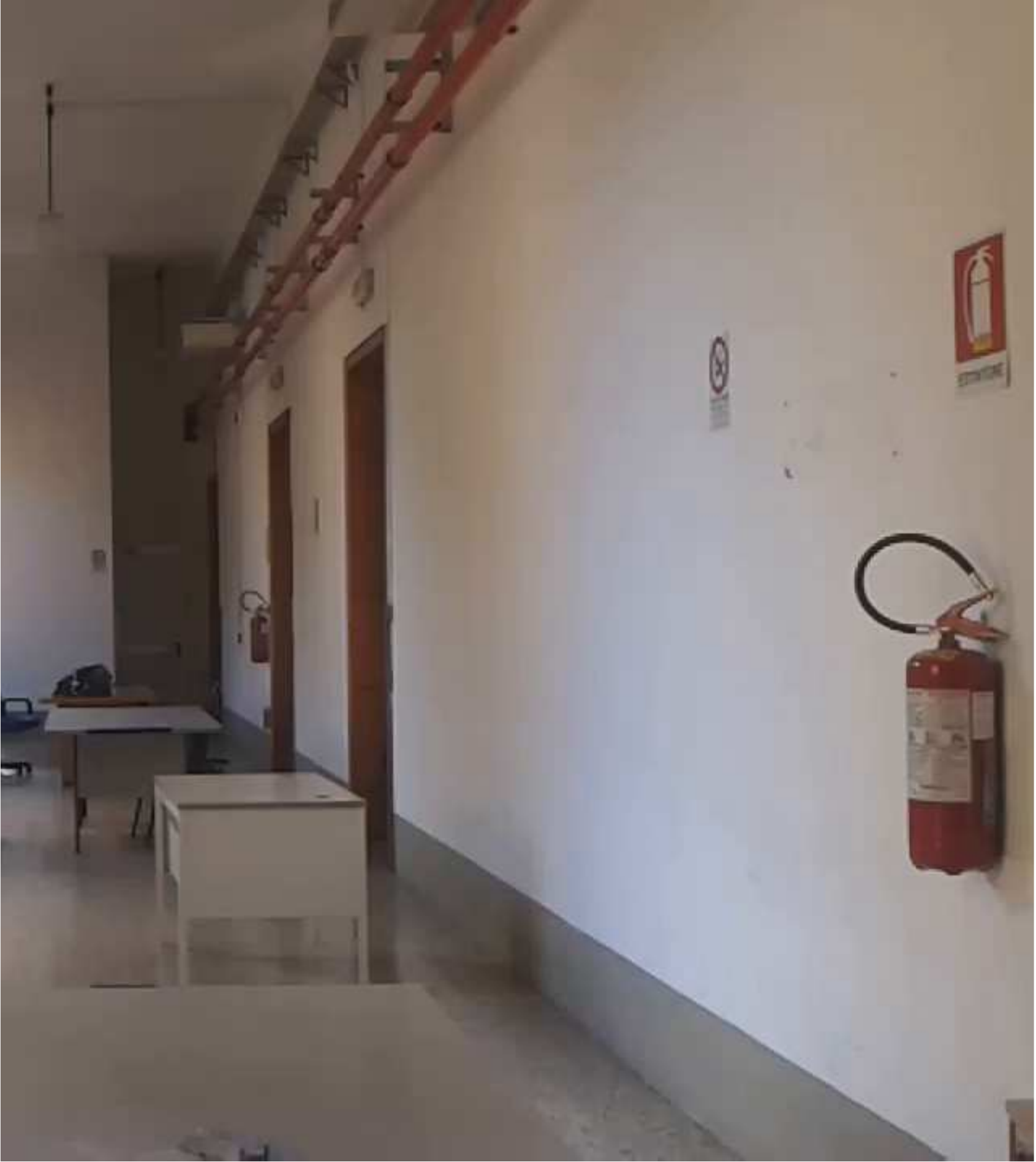}  & 
      \includegraphics[scale=0.15]{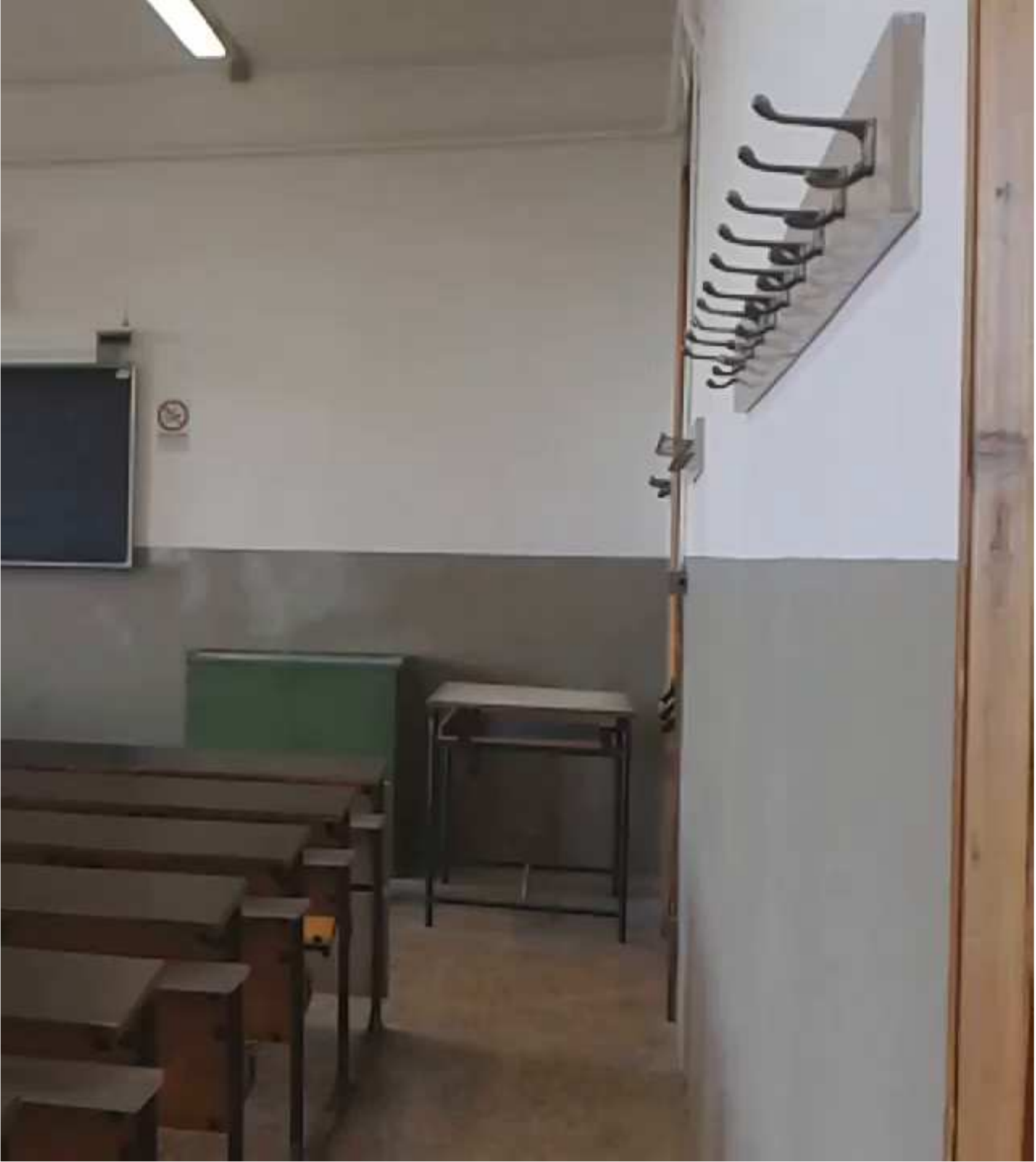} & 
      \includegraphics[scale=0.15]{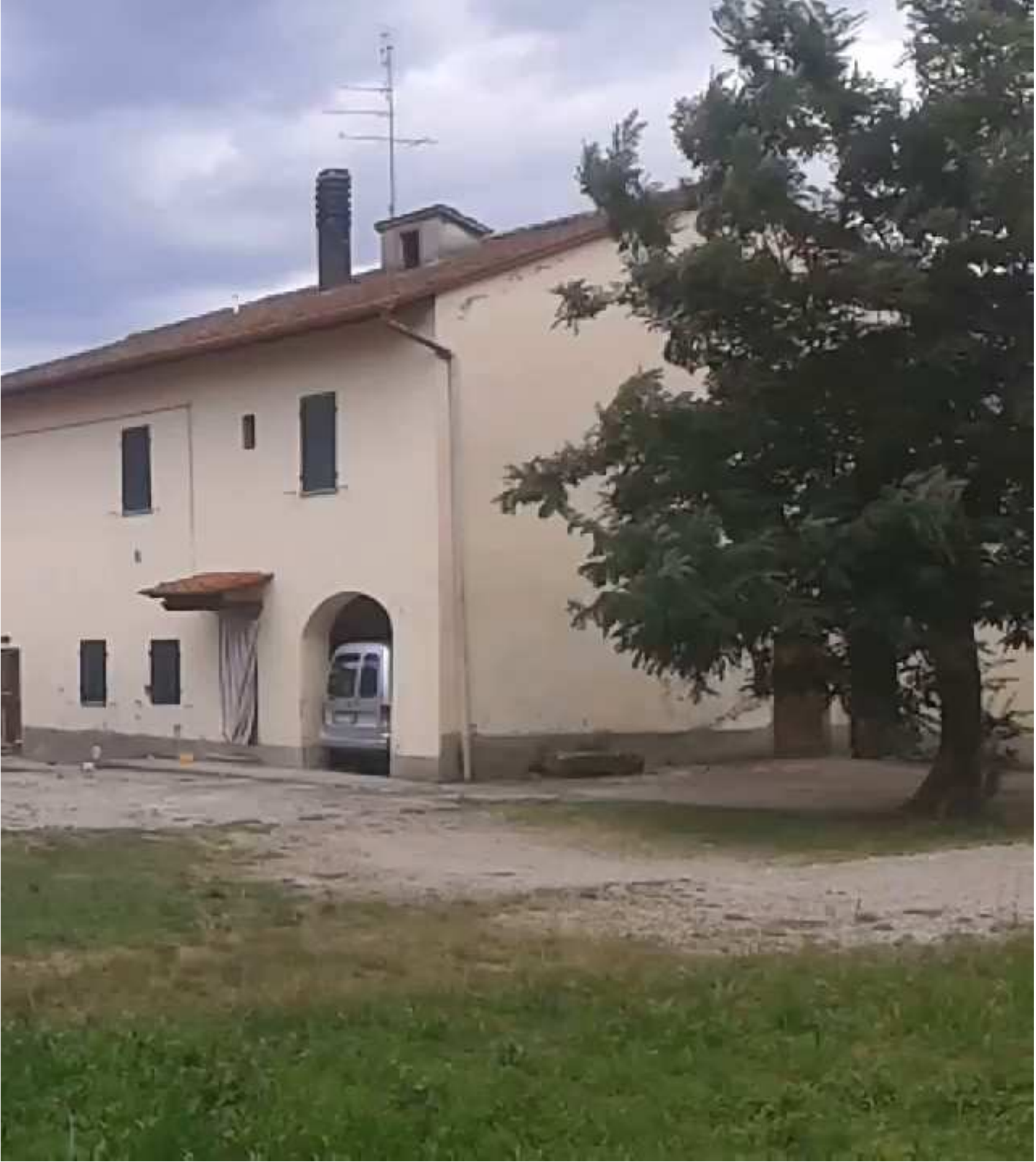}\vspace{-0.15cm} \\
      \includegraphics[scale=0.148]{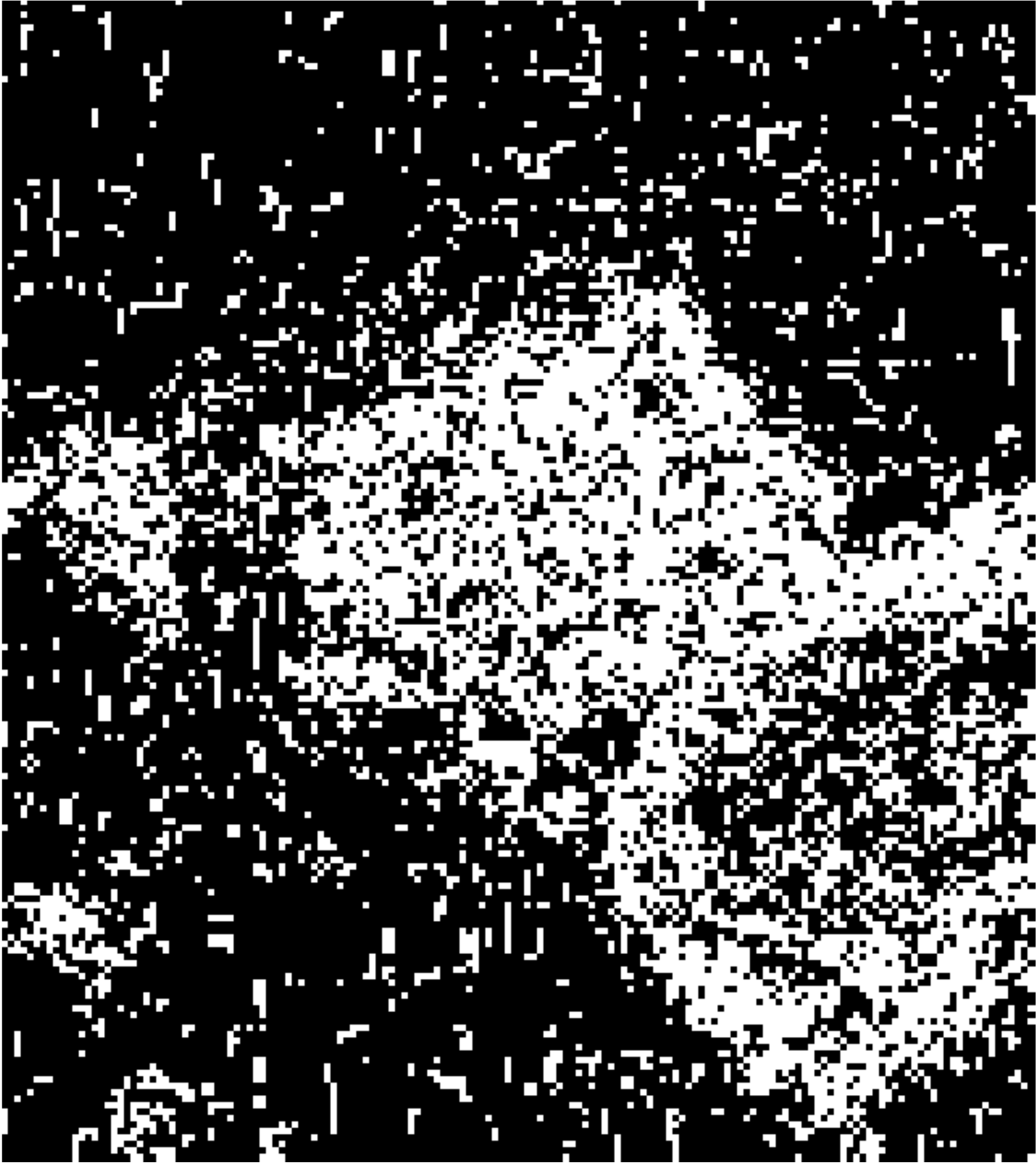}&  
      \includegraphics[scale=0.15]{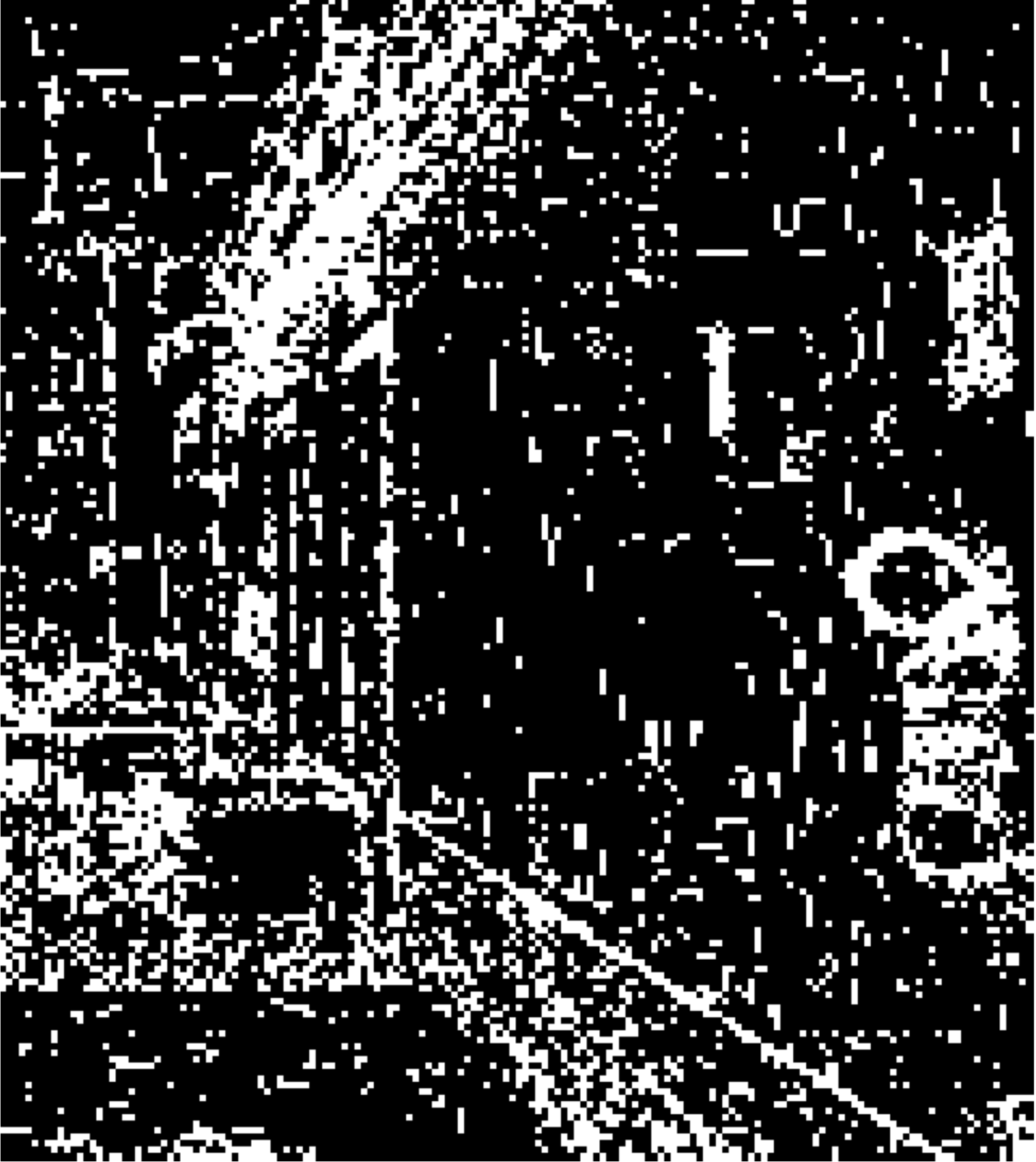} &
      \includegraphics[scale=0.15]{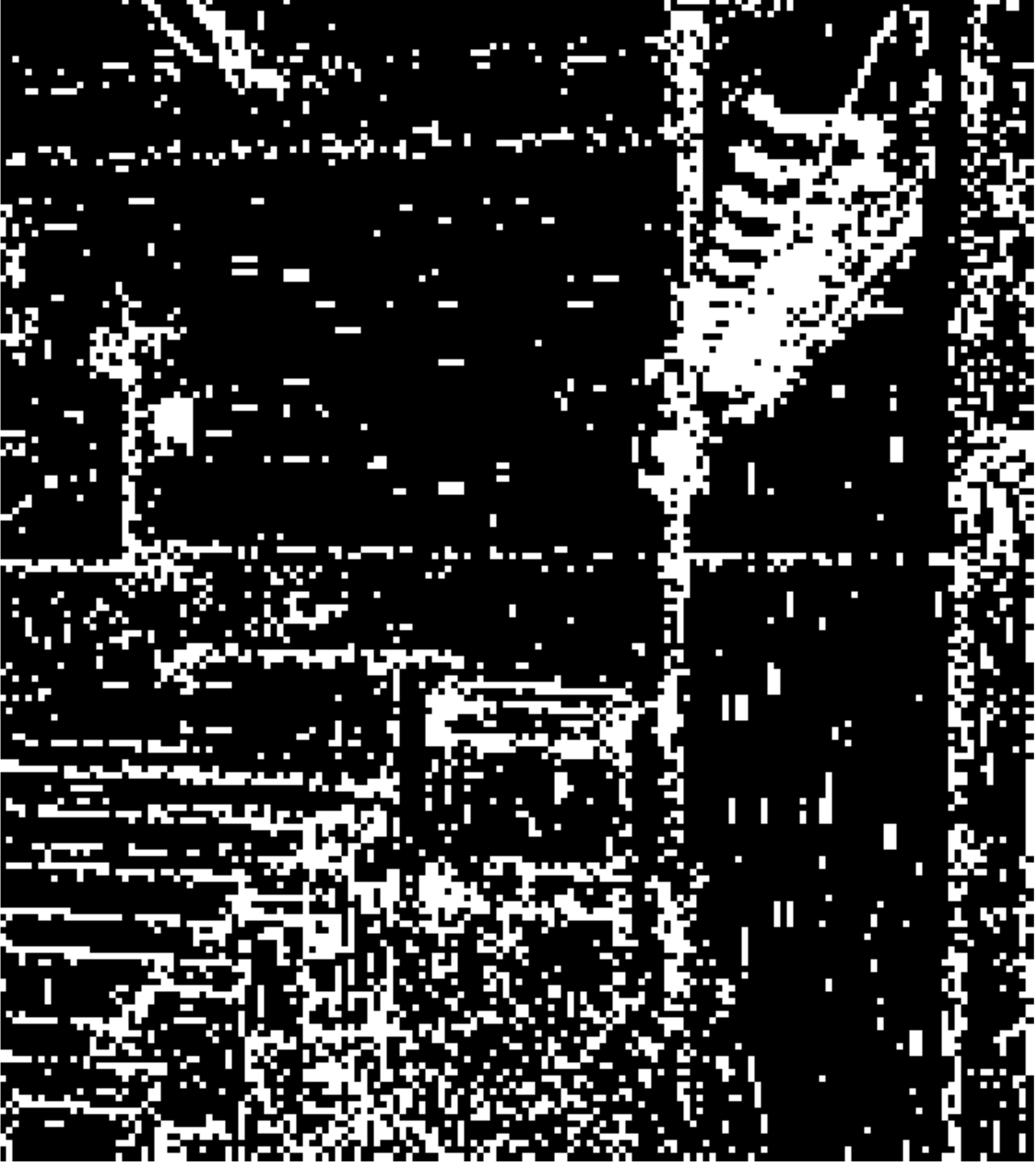} & 
      \includegraphics[scale=0.148]{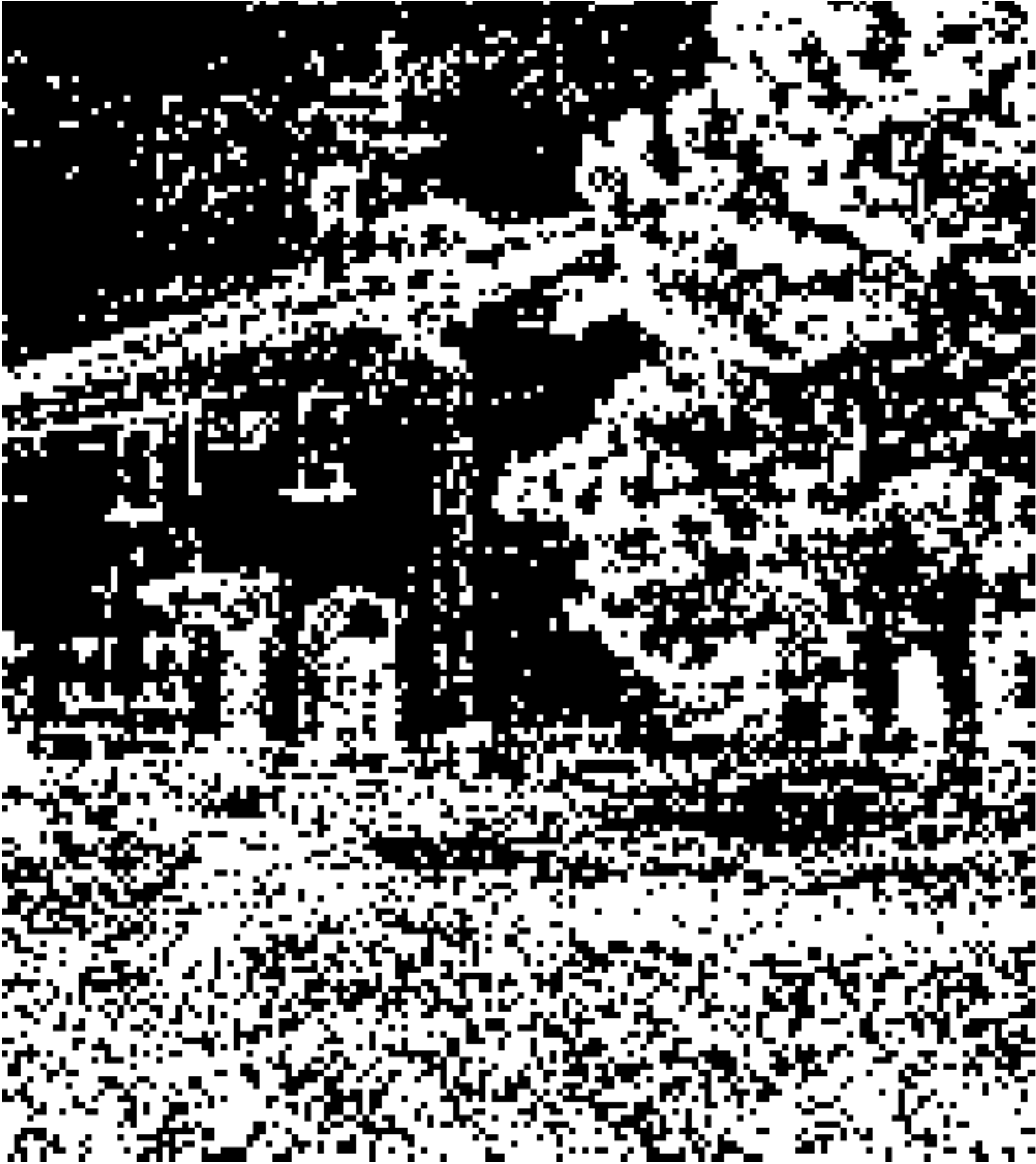} 
    \end{tabular}
%\end{table}
    \caption{I frames from YouTube videos (top) and their associated residual frame masks \textbf{M} (bottom)}
   \label{fig:frame mask sample}
\end{figure}

Fig. \ref{fig:frame mask sample} shows some examples of I frames from highly compressed (YouTube) videos and their associated frame masks $\mathbf{M}_k$. Only white regions (having a mask value of 1) in the frame mask will be used for fingerprint/video noise estimation. The figure shows that, due to intra-frame coding, most of the PRNU noise in uniform regions 
\textcolor{\revisionColor}{is replaced by the one in their prediction pixels}.

%\textcolor{\revisionColor}{is not valid}. 

In the block-based approach, the camera fingerprint $\mathbf{F}_v$ and the video noise $\mathbf{W}_v$ are computed with a modified version of the maximum likelihood PRNU fingerprint estimator (2) as provided in (\ref{eqn:K2 estimate}). 
\begin{equation}
\mathbf{F}_v = \frac{\sum\limits_{k=1}^l \mathbf{W}_{k}\mathbf{I}_{k}\mathbf{M}_k}{\sum\limits_{k=1}^{l}(\mathbf{I}_{k}\mathbf{M}_k)^{2} + \textrm{J}}
\label{eqn:K2 estimate}
\end{equation}
% Add comments to explain J : a matrix of ones which as the same dimensions with I and Mk
\textcolor{\revisionColor}{In (\ref{eqn:K2 estimate}), $l$ is the number of frames of the input video, $\mathbf{I}_{k}$ is the decoded image of the $k$ th video frame, $\mathbf{W}_{k}$ is the PRNU noise estimated from $\mathbf{I}_{k}$, and $\mathbf{M}_{k}$ is the frame mask. It should be noted that all the operations in (\ref{eqn:K2 estimate}) are element-wise. Different from (\ref{eqn:K estimate}), we add a J matrix (an all-one-matrix of the same size with the video resolution) to the equation's denominator to prevent a division by zero for the case where
$M_k(r,c)=0 \text{ , }\forall k$.}

%for the case the PRNU noise at $(r,c)$ location may be removed by another frame-block's noise, in all frames. In such a case, $M_k(r,c)=0 \text{ , }\forall k. $}

\section{Experimental Setup} \label{section : section experiment set}
To evaluate the efficiency of the proposed methods, we used a public dataset (VISION) \citep{Dasara2017} which comprises native and social media videos. In the dataset, there are 34427 images and 1914 videos acquired with 35 smart-phones from 11 major brands. 
In this paper, we only used videos acquired with devices which do not feature in-camera digital video stabilization (a subset of 19 cameras). 
For each device, there are three types of scenes: flat scenes (skies or flat walls), indoor scenes (classrooms, offices, halls, stores, etc.), and outdoor scenes (nature, garden, city, etc.). For each type of scene, three acquisition modes have been used to record videos: \textit{still} mode, where the user stands still while capturing the video; \textit{move} mode, where the user  walks while capturing the scene; and  \textit{pan} mode, where the user performs a pan and rotation while recording. 
All videos in the dataset have approximately the same duration of about 1 minute and 15 seconds, which corresponds approximately to 2000 frames per video. Table \ref{tab:listdevices} gives a list of devices used in the experiment (we keep the same IDs used in the dataset) and the number of native videos (with flat and natural content) for each device. By natural content, we mean outdoor or indoor scenes. Each native video has its YouTube version in the dataset. Thus, there are 354 native and 354 YouTube videos of the same scenes (see Table \ref{tab:listdevices}). 

\begin{center}
\begin{table}
%\begin{tabular}{|c|c|c|c|c|c|}
\footnotesize
\begin{tabular}{cccccccc}
%\toprule
\hline 
\hline 
ID & Resolution & Brand & Container & H.264 Profile & \#Flat&\#Natural& Total\\
\hline 
%\hline
D01 & 720p & S. Galaxy S3 Mini & MP4 & Baseline & 10 & 12 & 22\\
%\hline 
D03 & 1080p & Huawei P9  & MP4 & Constrained B.& 7 & 12 & 19 \\ 
%\hline 
D07 & 720p & Lenovo P70A  & 3GP & Baseline& 7 & 13 & 20 \\ 
%\hline 
D08 & 720p & S. Galaxy Tab 3 & MP4 & Constrained B. & 13 & 24 & 37 \\ 
%\hline 
D09 & 720p & A. iPhone 4  & MOV & Baseline & 7 & 12 & 19 \\ 
%\hline 
D11 & 1080p & S. Galaxy S3  & MP4 & Baseline & 7 & 12 & 19 \\ 
%\hline 
D13 & 720p & A. iPad 2  & MOV & Baseline & 4 & 12 & 16 \\ 
%\hline 
D16 & 1080p & Huawei P9 Lite & MP4 &Constrained B. & 7 & 12 & 19 \\ 
%\hline 
D17 & 1080p & M. Lumia 640 LTE  & MP4 & Main  & 4 & 6 & 10 \\ 
%\hline 
D21 & 1080p & Wiko Ridge 4G  & MP4 & Baseline & 4 & 7 & 11 \\ 
%\hline 
D22 & 720p & S. Galaxy Trend Plus & MP4 & Baseline & 4 & 12 & 16 \\ 
%\hline 
D24 & 1080p & X. Redmi Note 3  & MP4 &Baseline & 7 & 12 & 19 \\ 
%\hline 
D26 & 720p & S. Galaxy S3 Mini  & MP4 & Baseline & 4 & 12 & 16 \\ 
%\hline 
D27 & 1080p & S. Galaxy S5  & MP4 & High & 7 & 12 & 19 \\ 
%\hline 
D28 & 1080p & Huawei P8  & MP4 & Constrained B.& 7 & 12 & 19 \\ 
%\hline 
D30 & 1080p & Huawei Honor 5c  & MP4 & Constrained B.  & 7 & 12 & 19 \\ 
%\hline 
D31 & 1080p & S. Galaxy S4 Mini  & MP4 & High & 7 & 12 & 19 \\ 
%\hline 
D33 & 720p & Huawei Ascend  & MP4 & Constrained B. & 7 & 12 & 19 \\ 
%\hline 
D35 & 720p & S. Galaxy Tab A  & MP4 & Baseline & 4 & 12 & 16 \\ 
\hline 
%\hline
Total  & & &  &  &  124 & 230 & 354\\
%\hline
\hline 
\hline 
\end{tabular} 
\caption{The set of \emph{native} non-stabilized videos used in the experiments (by natural scenes we mean outdoor or indoor scenes).}
\label{tab:listdevices}
\end{table}
\end{center}

Tables \ref{tab: properties native videos} and \ref{tab:properties YT videos} present the average video bit-rates and average number of frames (I,B,P) of native and YouTube videos for each device in the VISION dataset, respectively. Videos uploaded to YouTube are re-encoded (re-compressed) to reduce their size. This can be noticed from the video bit-rates of native and YouTube videos. %We notice that a coding efficiency of up to ten times can be achieved with YouTube re-encoding. 
 When native videos are uploaded to YouTube, no re-scaling (down-sizing) is performed since YouTube supports 4K resolution, which is high enough for the videos in the VISION dataset. 
Videos that have $720p$ resolution are re-encoded by YouTube using the H.264 Main profile, while, $1080p$ videos are re-encoded using the High profile which has a bigger coding efficiency than the Main Profile. 
\textcolor{\revisionColor}{To achieve a high coding efficiency, the H.264 High profile coarsely quantizes the high-frequency DCT coefficients, whereas the Main profile performs a simple scalar quantization for all frequencies.}

All computations in our experiments was performed on a Dell Precision T3610 PC equipped with a 12 cored Xeon processor, 16 GB of RAM, running Ubuntu 16.04 LTS. Video frames were extracted and stored in an uncompressed format using \textit{ffmpeg} \citep{ffmpeg2017}. Video properties (type of frame, bit rate, resolution...) were extracted from videos using \textit{ffprobe}, included in \textit{ffmpeg} software. 

%% table giving the properties of the encoded videos according to the device
\begin{table}
\footnotesize
\begin{center}
\begin{tabular}{cccccccc}
\hline\hline
    ID  & Resolution & Container & \# Frames & \# I frames & \# P frames &\# B frames & Bit-rate (Kbps) \\ \hline 
    D01 & 720p       & MP4       &   2112 & 71 & 2041                 & 0                    & 3341           \\  
    D03 & 1080p      & MP4       &   2163 & 70                   & 2093                 & 0                    & 16567          \\  
    D07 & 720p       & 3GP       &   2125             & 71                   & 2054                 & 0                    & 8212           \\  
    D08 & 720p       & MP4       &   1912             & 64                   & 1847                 & 0                    & 11459          \\  
    D09 & 720p       & MOV       &   2027             &  46                  & 1567                 & 414                  & 6877           \\  
    D11 & 1080p      & MP4       &   2187             &  73                  & 2114                 & 0                    & 27466          \\  
    D13 & 720p       & MOV       &   2133             & 62                   & 1872                 & 198                  & 9407           \\  
    D16 & 1080p      & MP4       &   2155             & 70                   & 2085                 & 0                    & 16110          \\  
    D17 & 1080p      & MP4       &   2138             &  47                  & 879                  & 1211                 & 11056          \\
    D21 & 1080p      & MP4       &   2032             & 68                   & 1963                 & 0                    & 20004          \\  
    D22 & 720p       & MP4       &   2174             & 73                   & 2101                 & 0                    & 12025          \\  
    D24 & 1080p      & MP4       &   2031             &   68                 & 1963                 & 0                    & 19994          \\
    D26 & 720p       & MP4       &   2161             &   64                 & 1931                 & 167                  & 10901          \\  
    D27 & 1080p      & MP4       &   2088             & 70                   & 2017                 & 0                    & 17010          \\  
    D28 & 1080p      & MP4       &   2152             & 70                   & 2082                 & 0                    & 15942          \\  
    D30 & 1080p      & MP4       &   2162             &   70                 & 2092                 & 0                    & 17096          \\  
    D31 & 1080p      & MP4       &   2198             &  74                  & 2092                 & 0                    & 17005          \\  
    D33 & 720p       & MP4       &   2022             &  68                  & 1954                 & 0                    & 8032           \\  
    D35 & 720p       & MP4       &   2150             &  66                  & 1939                 & 145                  & 10903          \\ \hline
    \hline
\end{tabular}
\end{center}
\caption{Average video bit-rates and average number of frames (I,B,P) of \textit{native} videos per device in the VISION dataset}
\label{tab: properties native videos}
\end{table}
%% table giving the properties of the Youtube videos according to the device
\begin{table}
\footnotesize
\begin{center}
\begin{tabular}{ccccccccc}
\hline
\hline
%    \hline
    ID  & Resolution  & H.264 Profile& \# Frames & \# I frames & \# P frames & \# B frames & Bitrate (Kbps) \\ \hline
    D01 & 720p         &   Main  & 2113               & 22                   & 1366                 & 726                  & 1741           \\  
    D03 & 1080p        &  High   & 2161               & 52                   & 1520                 & 589                  & 3341           \\  
    D07 & 720p         &  Main   & 2153               & 42                   & 1307                 & 803                  & 1401           \\  
    D08 & 720p         &  Main   & 1920               & 31                   & 1216                 & 674                  & 1923           \\  
    D09 & 720p         &  Main   & 2115               & 50                   & 1675                 & 390                  & 1582           \\  
    D11 & 1080p        &  High   & 2187               & 22                   & 1177                 & 988                  & 3332           \\  
    D13 & 720p         &  Main   & 2128               & 35                   & 1416                 & 677                  & 1906           \\  
    D16 & 1080p        & High    & 2141               & 49                   & 1508                 & 584                  & 3176           \\  
    D17 & 1080p        &  High   & 2125               & 53                   & 1411                 & 661                  & 3654           \\  
    D21 & 1080p        &  High   & 2174               & 97                   & 1511                 & 565                  & 3050           \\  
    D22 & 720p         &  Main   & 2172               & 27                   & 1415                 & 730                  & 1997           \\  
    D24 & 1080p        &   High   & 2029               & 39                   & 1228                 & 763                  & 3228           \\  
    D26 & 720p         &   Main  & 2156               & 24                   & 1421                 & 712                  & 1973           \\  
    D27 & 1080p        &  High   & 2088               & 46                   & 1429                 & 613                  & 3210           \\  
    D28 & 1080p        &  High   & 2152               & 21                   & 1288                 & 843                  & 3835           \\  
    D30 & 1080p        &  High   & 2151               & 34                   & 1339                 & 779                  &  3608          \\  
    D31 & 1080p        &  High    & 2198               & 31                   & 1296                 & 871                  & 3668           \\  
    D33 & 720p         &  Main   & 2080               & 20                   & 1568                 & 492                  & 1988           \\  
    D35 & 720p         &  Main   & 2155               & 29                   & 1359                 & 767                  & 1876           \\ \hline
    \hline
    \end{tabular}
\end{center}
\caption{Average video bit-rates and average number of frames (I,B,P) of YouTube videos per device in the VISION dataset}
\label{tab:properties YT videos}
\end{table}

%%%%%%%%%%%%%%%%%%%%%%%%%%%%%%%%%%%%%%%%%%%%%%%%%%%%%%%%%%%%%%%%%%%%%%%%%%%%%%%%%%%%%%%%%%%%%%%%%%%%%%

\section{Experimental Results}
In this section, we evaluate the efficiency of the frame-based video source attribution (which is the one commonly used in the literature) and the block-based approach (the one we propose) under six different scenarios with an increasing level of difficulty as given in Table \ref{tab:Test scenarios}. Scenario 1 and Scenario 3 correspond to cases where the suspect cameras are available at hand. Thus, native flat content videos can be recorded for camera fingerprint estimation. In the remaining scenarios, neither any suspect camera nor native-flat videos are available. \textcolor{\revisionColor}{In all the scenarios, we want to figure out whether two input videos (query and reference) originate from the same device or not.}

For each scenario, all reference videos are matched against all query videos of the same device and other devices having the same resolution with video PRNU fingerprints. Thus, for each device, there are much more non-matching videos than matching videos. It should be noted that the video PRNU fingerprint estimation is performed in the same way both for the reference and the query videos.

\begin{table}
\footnotesize
\center
\begin{tabular}{ccc}
\hline
\hline
Scenario & Reference video & Query video  \\
\midrule
Scenario-1 & Native-flat & Native-natural \\
Scenario-2 & Native-natural & Native-natural  \\
Scenario-3 & Native-flat & YouTube-natural \\
Scenario-4 & Native-natural & YouTube-natural \\
Scenario-5 & YouTube-flat & YouTube-natural \\
Scenario-6 & YouTube-natural & YouTube-natural\\
\hline
\hline
\end{tabular}
\caption{Test scenarios used to evaluate the efficiency of the \emph{frame-based} and the \emph{block-based} methods}
\label{tab:Test scenarios}
\end{table}

\subsection{Source device attribution of native videos}
Here, we consider the case where we perform source device attribution exclusively on native videos. This corresponds to Scenario 1 and Scenario 2. Table \ref{Tab:Global AUC native videos} gives the overall source attribution accuracy (considering all the devices) of each approach on 720p and 1080p videos. The accuracy is measured using Area Under the Curve (AUC) of Receiver Operating Characteristic (ROC) computed from PCE distributions of matching and non-matching cases for all cameras listed in Table \ref{tab:listdevices}. Table \ref{Tab:Global AUC native videos} shows that, when performing source device attribution on native videos, an accuracy of almost 100\% can be achieved by using only I frames. This conforms with the results obtained in \citep{Samet2016}. The same results are obtained when using all video frames or the block-based method. 
Nevertheless, using only I frames \textcolor{\revisionColor}{ (C$_{1}$) is the best option if we are concerned with the computation time}.

%Command to create continuous line in multicolumn tables
\aboverulesep=0ex
\belowrulesep=0ex
\renewcommand{\arraystretch}{1.1}
%%%%%%%%%%%%%%%%%%%%%%%%%%%%%%%%%%%%%%%%%%%%%%%%%%%%%%%%%%%%%%
\begin{table}[]
\centering
\footnotesize
\begin{center}
\begin{tabularx}{1.0\textwidth}{cCCCC}
\hline
\hline
 & \multicolumn{2}{c}{ Flat vs. Natural}  & \multicolumn{2}{c}{ Natural vs. Natural} \\
            %\midrule
  Methods             & 720p                   & 1080p                   & 720p                     & 1080p                    \\
            \midrule
I-frames & 1.00                   & 0.99                    & 0.99                     & 1.00                     \\
All-frames  & 1.00                   & 1.00                    & 0.99                     & 0.99                     \\
Proposed method    & 1.00                   & 1.00                    & 1.00                     & 1.00                    \\
\hline
\hline

\end{tabularx}
\end{center}

\caption{AUC values for source device attribution of \emph{native} videos}
\label{Tab:Global AUC native videos}
\end{table}

\subsection{Source device attribution of YouTube videos with native reference videos}
\begin{figure}[]
\begin{center}
%    \centering
       \subfloat[720p videos]{{\includegraphics[scale=0.5]{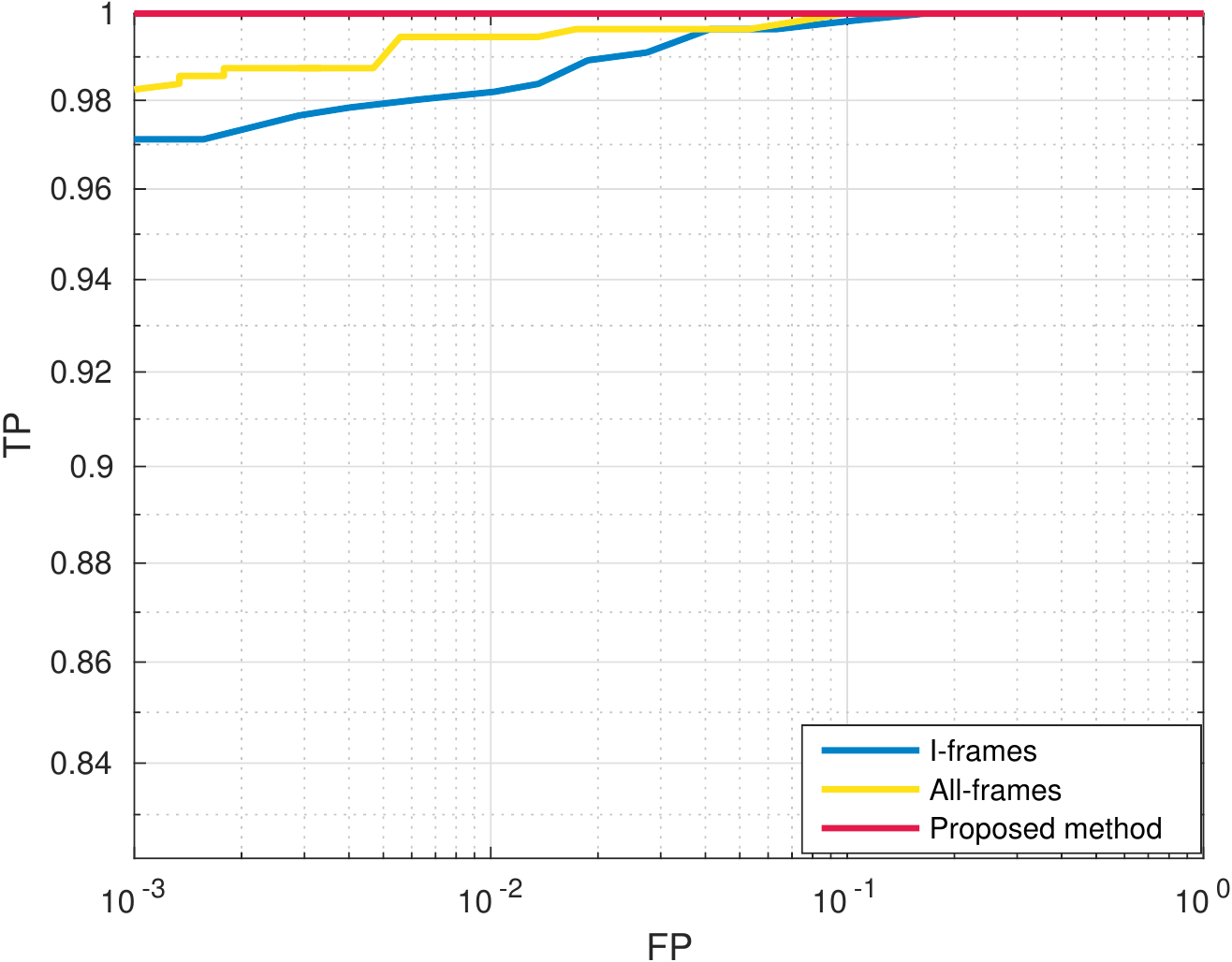}}}
       \subfloat[1080p]{{\includegraphics[scale=0.5] {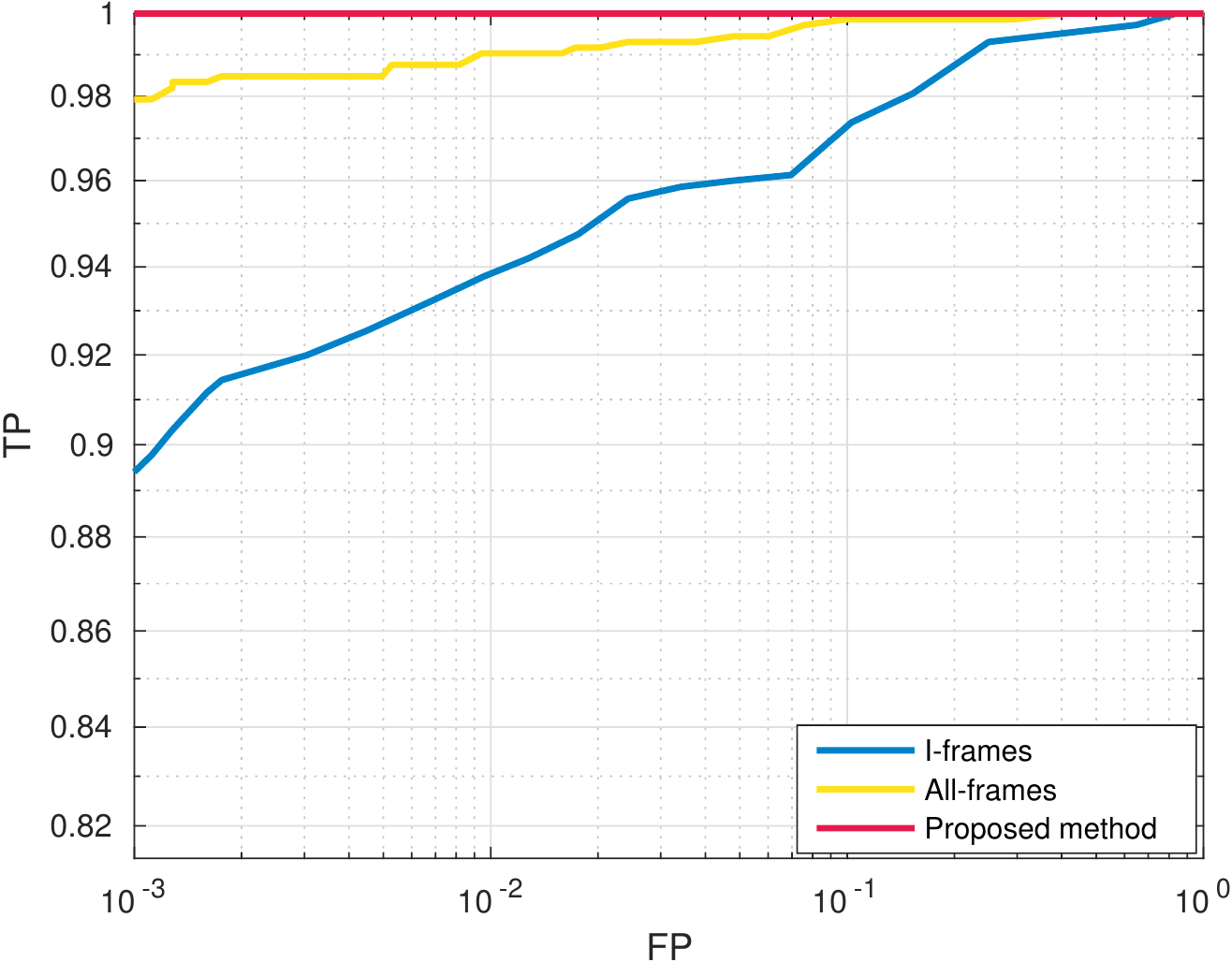}}}
\caption{ROC curves for native (flat) vs. YouTube (natural) videos matching}
\label{fig:ROC curves Youtube videos with native flat reference videos}
\end{center}
\end{figure}

\begin{figure}[]
    \centering
       \subfloat[720p videos]{{\includegraphics[scale=0.5]{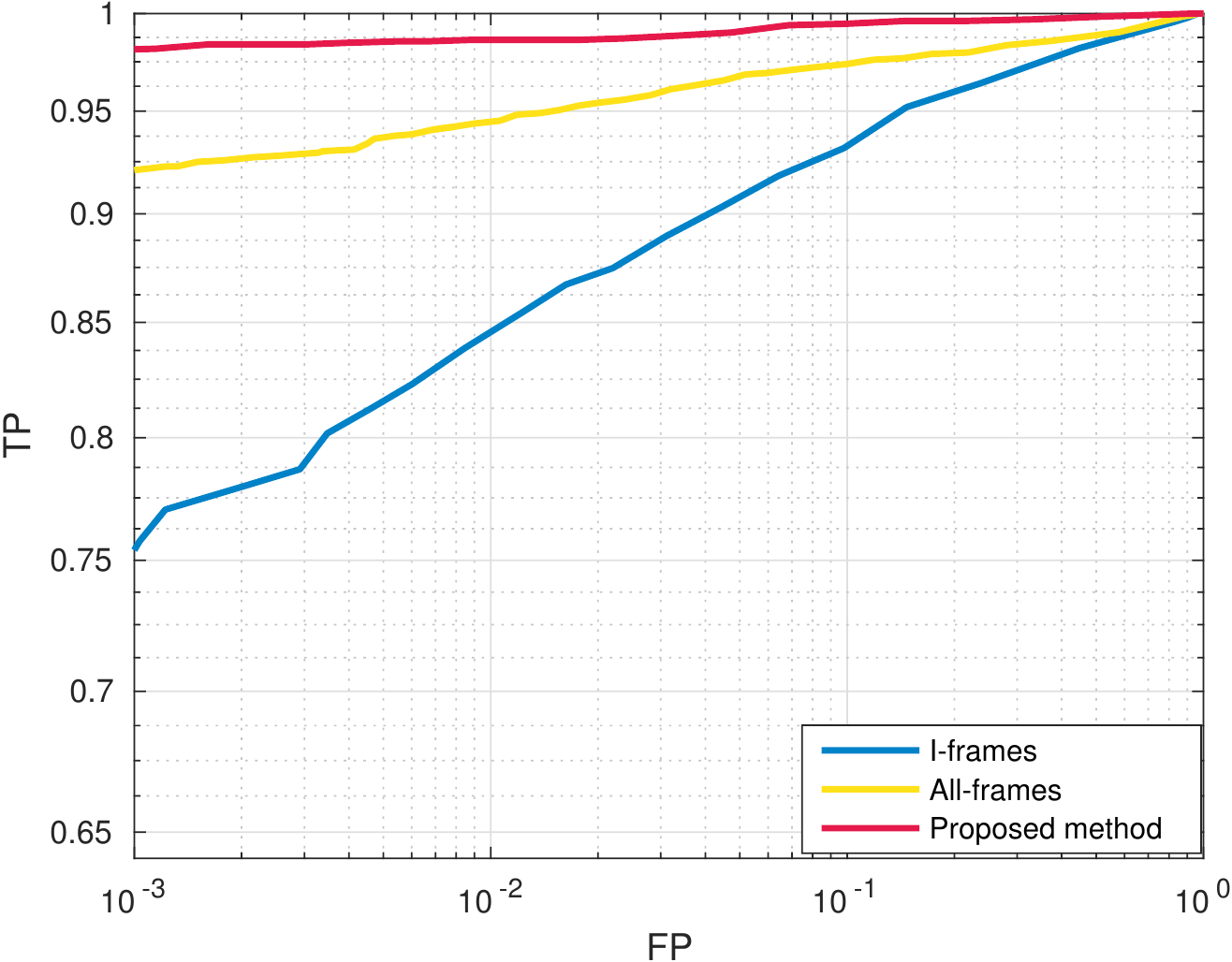}}}
       \subfloat[1080p]{{\includegraphics[scale=0.5] {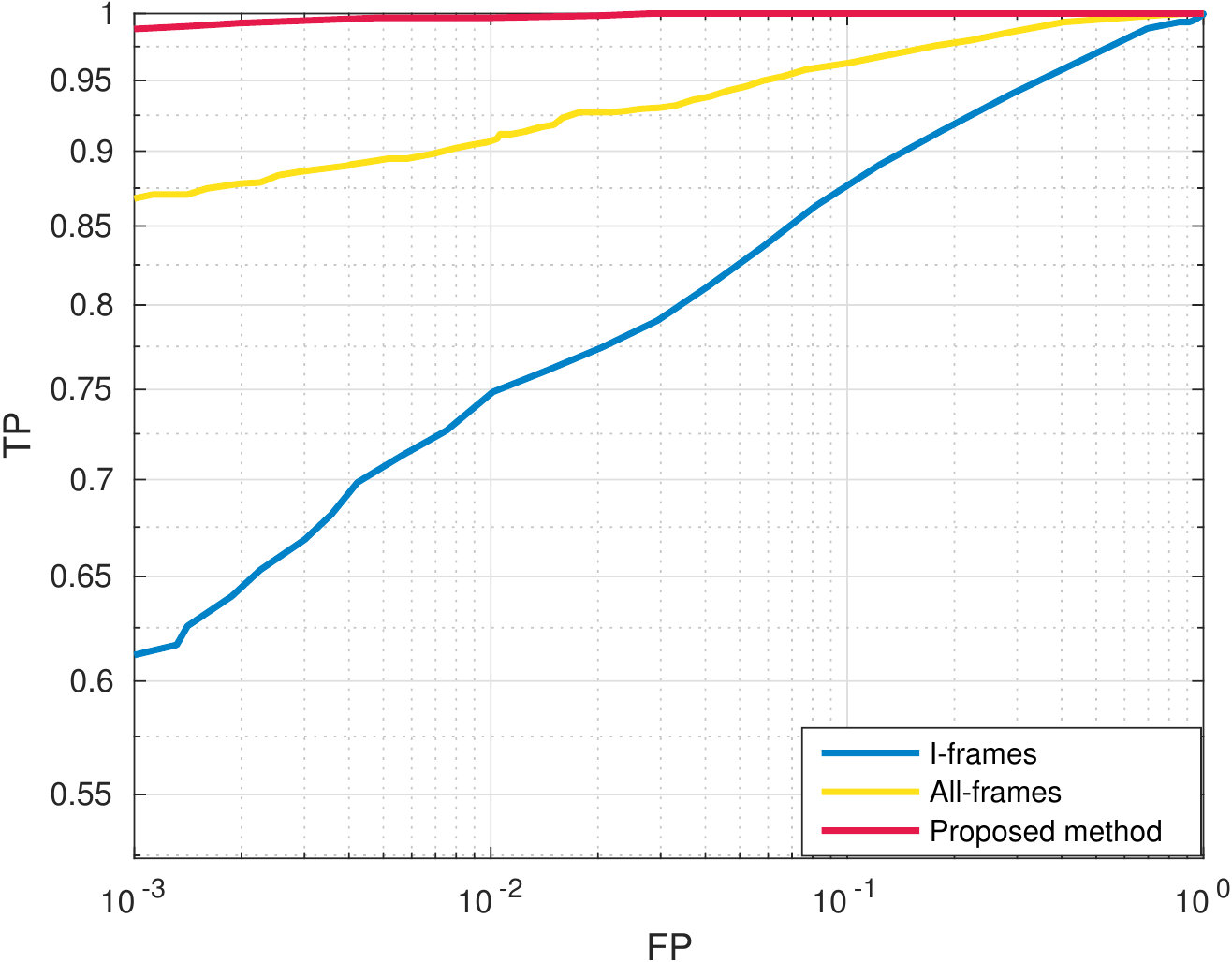}}}
\centering
\caption{ROC curves for native (natural) vs. YouTube (natural) videos matching}
\label{fig:ROC curves Youtube videos with native natural reference videos}
\end{figure}

Here, we consider Scenarios 3 and 4, where reference videos are native videos (flat or natural content) meanwhile query videos are YouTube natural-content videos. Table \ref{tab:AUC values per device native-YouTube videos} gives the AUC values per device for the 720p and 1080p video sets, Fig. \ref{fig:ROC curves Youtube videos with native flat reference videos} and Fig. \ref{fig:ROC curves Youtube videos with native natural reference videos} give the associated ROC curves. Globally, a high accuracy is achieved by both methods, but the block-based approach slightly surpasses the frame-based approach. 

\subsection{Source device attribution of YouTube videos with YouTube reference videos}
\begin{figure}[]
    \centering
       \subfloat[720p videos]{{\includegraphics[scale=0.5]{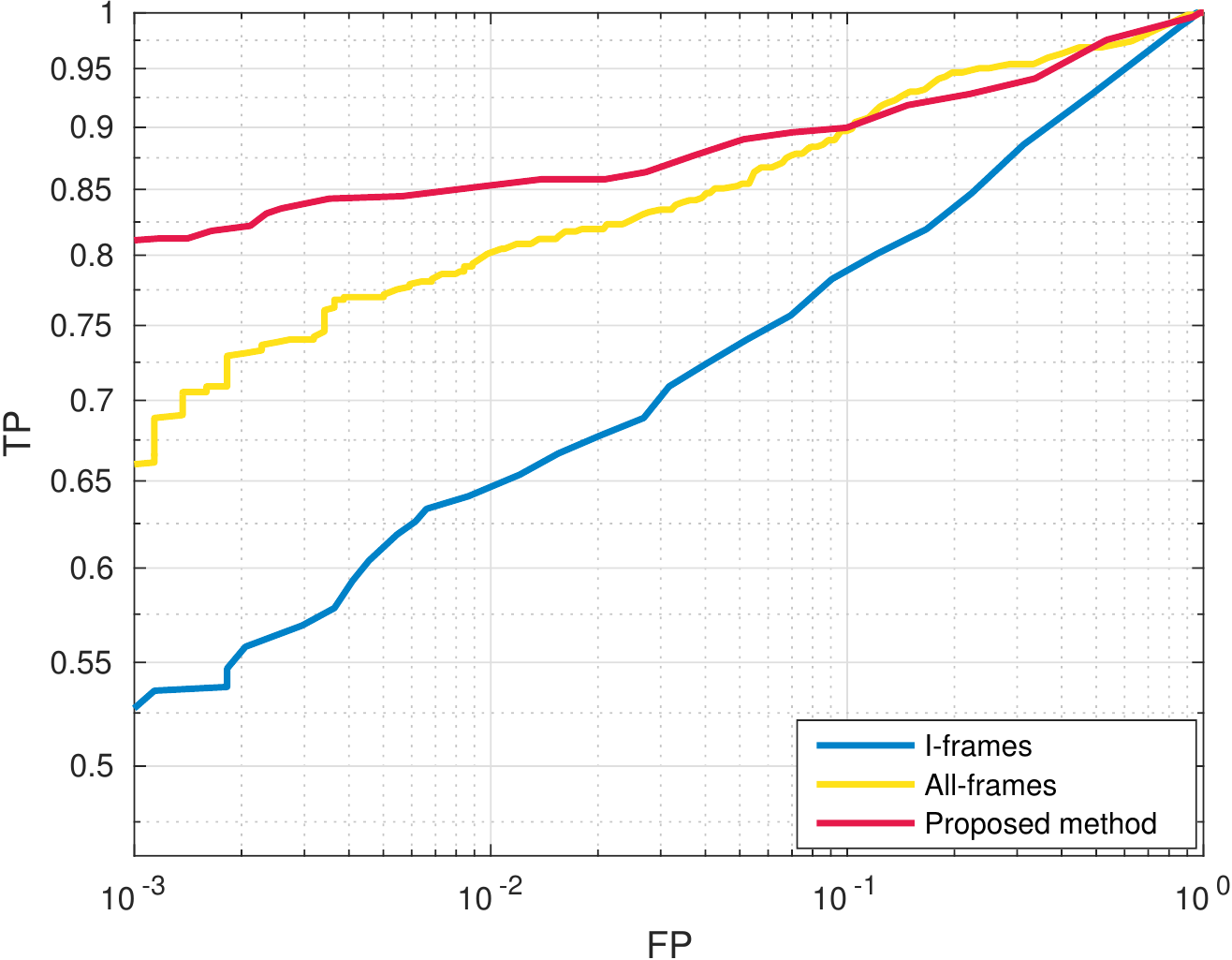}}}
       \subfloat[1080p]{{\includegraphics[scale=0.5] {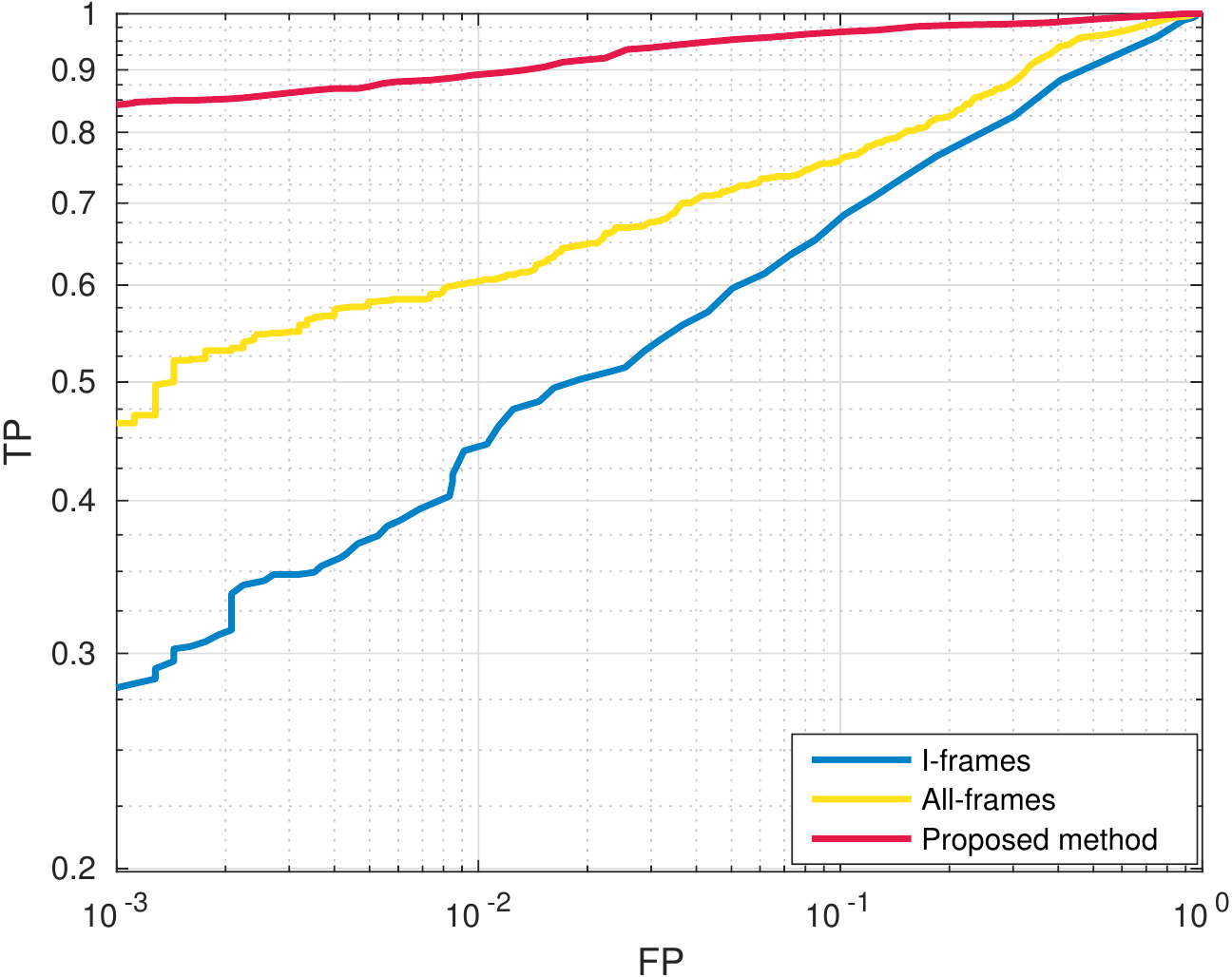}}}
\caption{ROC curves for YouTube (flat) vs. YouTube (natural) videos matching}
\label{fig:ROC curves Youtube videos with Youtube flat reference videos}
\end{figure}

\begin{figure}[]
    \centering
       \subfloat[720p videos]{{\includegraphics[scale=0.51]{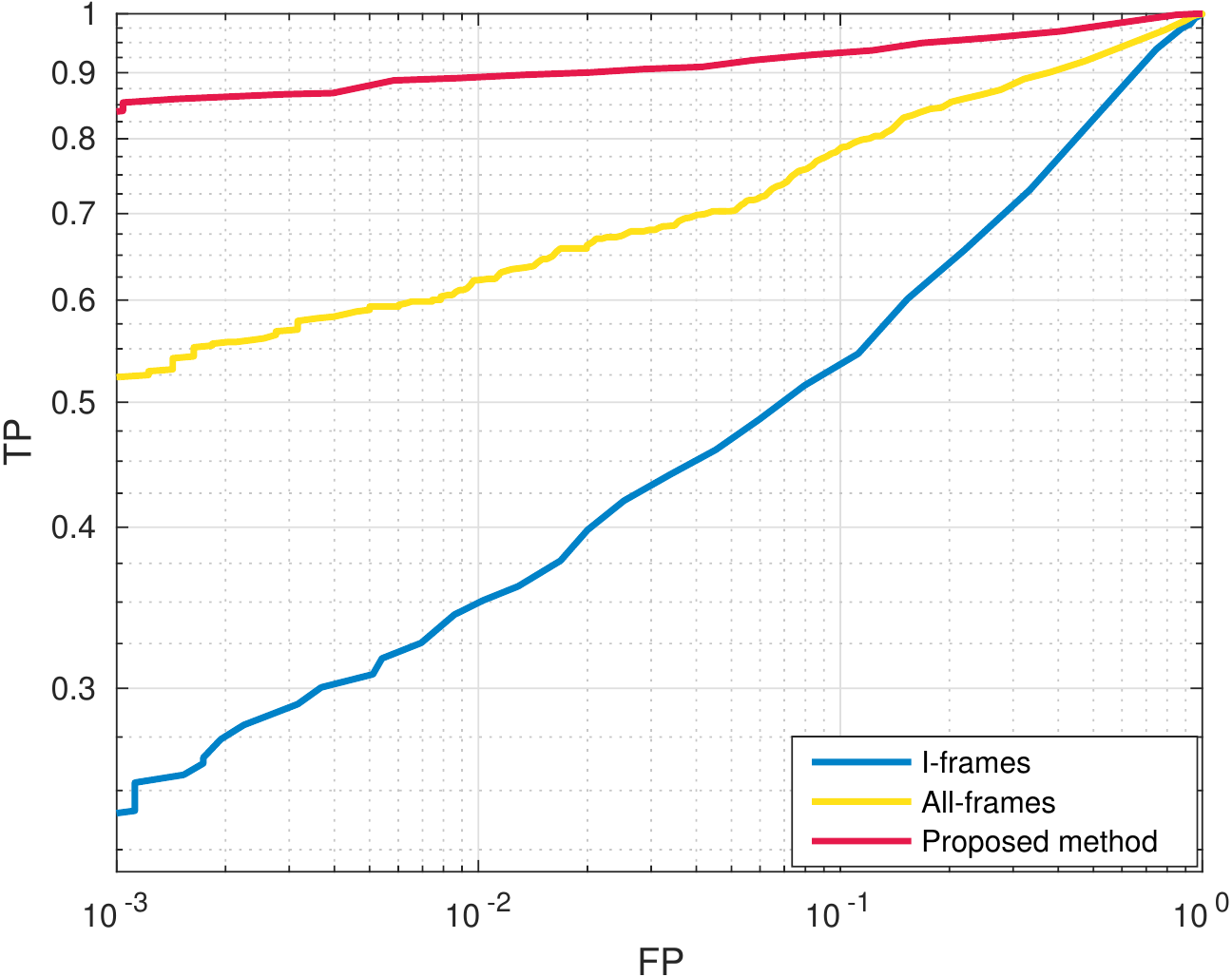}}}
       \subfloat[1080p]{{\includegraphics[scale=0.45] {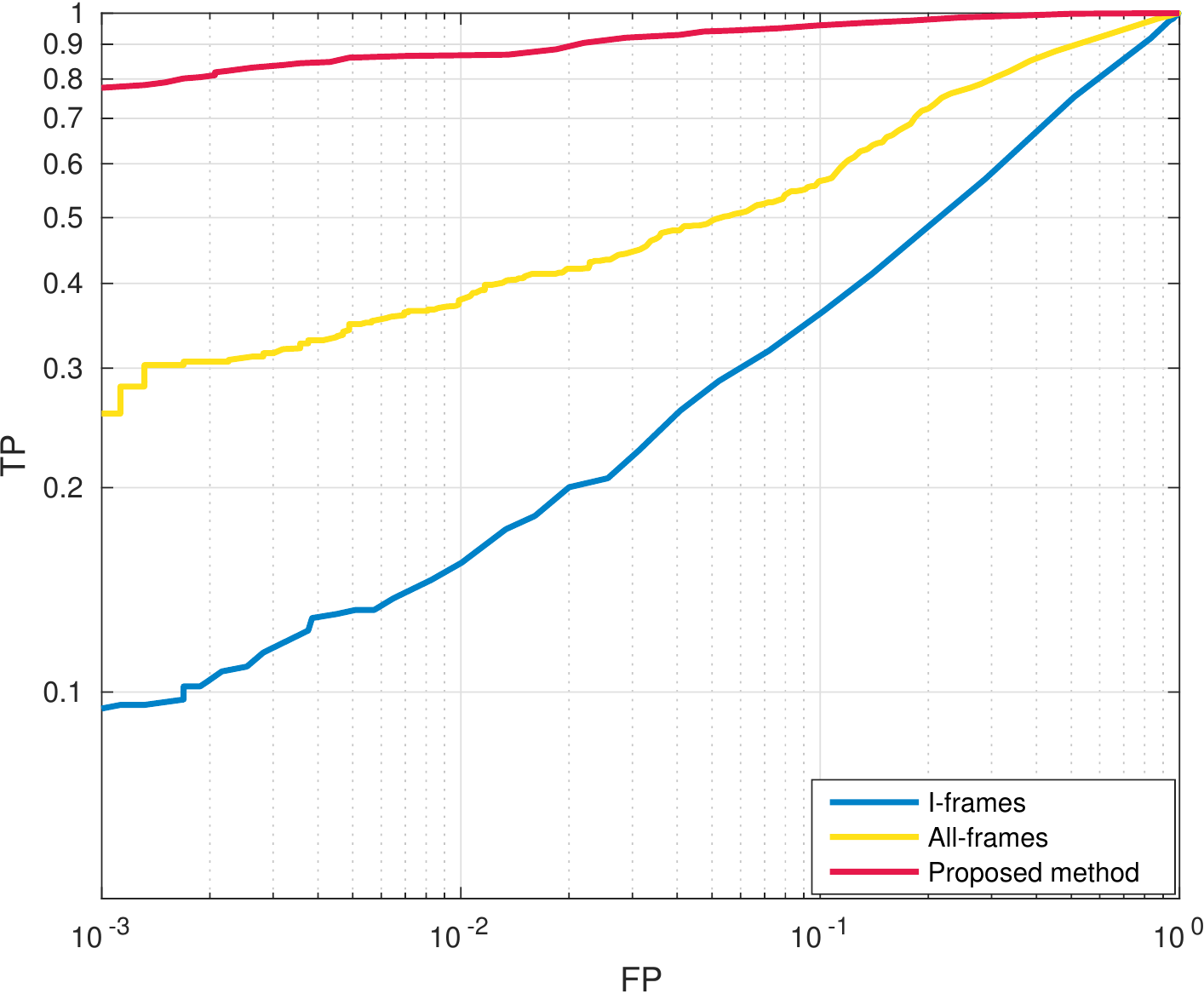}}}
\caption{ROC curves for YouTube (natural) vs. YouTube (natural) videos matching}
\label{fig:ROC curves Youtube videos with Youtube natural reference videos}
\end{figure}

We now consider the case where the reference and the query videos are all YouTube videos (Scenarios 5 and 6). This is the most challenging case since both reference and query videos are heavily compressed. Table \ref{tab:AUC values per device YouTube-YouTube videos} gives the global AUC values of these two Scenarios for all the devices in the set. Corresponding ROC curves are depicted in Fig. \ref{fig:ROC curves Youtube videos with Youtube flat reference videos} and Fig. \ref{fig:ROC curves Youtube videos with Youtube natural reference videos}. These results show the effectiveness of the block-based approach that is capable to link YouTube videos with a higher accuracy than the frame-based methods. 

\textcolor{\revisionColor}{
\subsection{The impact of video motion on source device attribution}
To evaluate the impact of video motion on the accuracy of source device attribution, we consider the Scenario-6 (YouTube-natural vs. YouTube natural videos matching) with two video classes: \textit{still} videos, and \textit{move} videos. The still-class refers to videos recorded without significant camera movement and content change. The move-class refers to videos recorded with camera motion. This experiment was done using all the cameras in Table \ref{tab:listdevices} which corresponds to 33  \emph{still} 720p videos, 27 \emph{still} 1080p videos, 33 \emph{move} 720p videos, and 27 \emph{move} 1080p videos. 
Table \ref{tab: AUC with regard to video motion} gives AUC values of the frame-based and the block-based methods on these two video classes. The table shows that, when the block-based method is used, the highest source attribution accuracy is always obtained when the query videos have significant amount of motion compared to the still case. The same remark holds for the videos encoded with the H.264 High profile ({1080p YouTube videos}) for both the frame-based and the block-based methods. 
This is due to the fact that, when a video has lot of motion (scene change), the encoder has to keep more non-zero DCT-AC coefficients to cope with the changes in the scene. We also notice from the table that I frames of {720p} videos acquired with the still mode have more PRNU noise than their move counterpart.
}
%The block-based approach brings an improvement of up to 15\% to the accuracy (AUC) of the frame-based approach. 
%Regarding True Positive classification rate (TP), an improvement of up to 40\% is noticed.
%%%%%%%%%%%%%%%%%%%%%%%%%%%%%%%%%%%%%%%%%%%%%%%%%%%%%%%%%%%%%%%%%%%%%%%%%%%%%%%%%%%
\begin{table}[]
\centering
\footnotesize
\begin{center}
\begin{tabularx}{1.1\textwidth}{ccccc}
\hline
\hline
 & \multicolumn{2}{c}{Native (flat) vs. YouTube (natural)}  & \multicolumn{2}{c}{ Native (natural) vs. YouTube (natural)} \\
            
  Methods             & 720p                   & 1080p                   & 720p                     & 1080p \\
            \midrule
I-frames        &0.99 &0.99 &0.97 & 0.94 \\
All-frames      &0.99 &0.99 &0.98 & 0.98\\
Proposed method &1.00 &1.00 &0.99 & 1.00\\
\hline
\hline
\end{tabularx}
\end{center}
\caption{AUC values for source attribution of \emph{YouTube} videos with \emph{native} reference videos}
\label{tab:AUC values per device native-YouTube videos}
\end{table}
%%%%%%%%%%%%%%%%%%%%%%%%%%%%%%%%%%%%%%%%%%%%%%%%%%%%%%%%%%%%%%%%%%%%%%%%%%%%%%%%

%%%%%%%%%%%%%%%%%%%%%%%%%%%%%%%%%%%%%%%%%%%%%%%%%%%%%%%%%%%%%%%%%%%%%%%%%%%%%%%%
\begin{table}[]
\centering
\footnotesize
\begin{center}
\begin{tabularx}{1.15\textwidth}{ccccc}
\hline
\hline
 & \multicolumn{2}{c}{YouTube (flat) vs. YouTube (natural)}  & \multicolumn{2}{c}{YouTube (natural) vs. YouTube (natural)} \\
            
  Methods             & 720p                   & 1080p                   & 720p                     & 1080p \\
            \midrule
I-frames        &0.90 &0.86 &0.78 &0.69 \\
All-frames      &0.95 &0.91 &0.89 &0.83 \\
Proposed method &0.95 &0.98 &0.97 &0.98 \\
\hline
\hline
\end{tabularx}
\end{center}
\caption{AUC values for source attribution of \emph{YouTube} videos with \emph{YouTube} reference videos}
\label{tab:AUC values per device YouTube-YouTube videos}
\end{table}
%%%%%%%%% table to show the impact of video motion on source device identification %%%%%%%%%%%%%%%%%%%%%%%%%%%%%%%%%%%%%%%%%%%%%%%%%%%%%%%%%%

\begin{table}[]
\centering
\footnotesize
\begin{center}
\begin{tabular}{ccccc}
\hline
\hline
& \multicolumn{2}{c}{(still vs. still)}  & \multicolumn{2}{c}{(move vs. move)} \\
  Methods             & 720p                   & 1080p                   & 720p                     & 1080p \\
            \midrule
I-frames        &0.83 &0.63 &0.69 &0.63  \\
All-frames      &0.87 &0.74 &0.86 &0.87 \\
Proposed method       &0.94 &0.95 &0.98 &0.98 \\
\hline
\hline
\end{tabular}
\end{center}
\caption{\textcolor{\revisionColor}{
AUC values with regard to motion in  \emph{YouTube-natural} videos}}
\label{tab: AUC with regard to video motion}
\end{table}

%%%%%%%%%%%%%%%%%%%%%%%%%%%%%%%%%%%%%%%%%%%%%%%%%%%%%%%%%%%%%%%%%%%%%%%%%

%%%%%%%%%%%%%%%%%%%%%% Figures removed in the final version %%%%%%%%%%%%%%%%%
%\begin{figure}[H]
%\begin{center}
%%    \centering
%       \subfloat[720p videos]{{\includegraphics[width=6.5cm, %height=5.5cm]{./YT_YT_move_move_720p.pdf}}}
%       \subfloat[1080p]{{\includegraphics[width=6.5cm, height=5.5cm] {./YT_YT_move_move_1080p.pdf}}}
%\caption{\textcolor{\revisionColor}{ROC curves for YouTube (natural-move) %vs. YouTube (natural-move) videos matching}}
%\label{fig:ROC curves Youtube natural-move vs Youtube natural-move %matching}
%\end{center}
%\end{figure}
%\begin{figure}[H]
%\begin{center}
%%    \centering
%       \subfloat[720p %videos]{{\includegraphics[scale=0.6]{./YT_YT_still_still_720p.pdf}}}
%       \subfloat[1080p]{{\includegraphics[scale=0.6] %{./YT_YT_still_still_1080p.pdf}}}
%\caption{\textcolor{\revisionColor}{ROC curves for YouTube %(natural-still) vs. YouTube (natural-still) videos matching}}
%\label{fig:ROC curves Youtube natural-still vs Youtube natural-still %matching}
%\end{center}
%\end{figure}
%%%%%%%%%%%%%%%%%%%%%%%%%%%%%%%%%%%%%%%%%%%%%%%%%%%%%%%%%%%%%%%%%%
\section{Discussion and Conclusion}
%In this paper, a novel approach to PRNU-based source device attribution for higly compressed digital videos is proposed. 
In this study, we investigate the problem of source verification of any two videos (query and reference) to determine whether they originate from the same camera device or not. The proposed scheme takes into account the effect of H.264/AVC video encoding on the PRNU noise in video frames. We first determine a necessary condition for the PRNU noise to survive H.264/AVC compression; then, propose a modified maximum likelihood estimator of video PRNU noise/fingerprint. 

The efficacy of the proposed method (called block-based method) is evaluated on non-stabilized videos in the VISION database with Receiver Operating Characteristic (ROC) plots and Area-Under-the-Curve (AUC) measurements. The proposed method is compared with existing video PRNU fingerprint estimation methods using all frames (I, B, P) and solely I-frames under different scenarios based on video content (flat/natural) and video encoding (native/YouTube).

%It has experimentally been shown on a large set of videos, that, an accurate source attribution of native videos taken with today's smart-phone and tablets cameras can be achieved using only I frames from videos of around 80 seconds length. 
%The devices used in the experiments implement different H.264/AVC profiles (Baseline, Main, and High) thus, achieve different levels of coding efficiency. 

From experimental results, we can infer that an accurate source identification of YouTube videos can be achieved using frame-based methods if we have a fingerprint estimated from I frames of a flat content native video (Table 7).
Oppositely to what has been stated in previous studies in the literature, using all frames (I, B, and P frames) yields better source attribution accuracy than using only I frames even when the investigated videos have been re-compressed by YouTube. 
\textcolor{\revisionColor}{This means that, despite of compression, there is still enough valid PRNU noise in P and B frames to improve the I-frames estimated PRNU fingerprint. Figures \ref{fig:ROC curves Youtube videos with native flat reference videos} to \ref{fig:ROC curves Youtube videos with Youtube natural reference videos} show that, when the frame-based methods are used, the best attribution accuracy is obtained when reference PRNU fingerprints are estimated from flat-content videos.
 Also, we notice from these figures that, when the frame-based approach is used, the attribution accuracy of {720p} videos is always higher than the one of {1080p} videos (which is contradictory to what one could have expected since higher resolution implies more PRNU information). This is because YouTube re-compresses {1080p} videos using the H.264 High profile (whereas it uses the Main profile for {720p} videos) yielding a higher degradation of the PRNU noise in the encoded blocks. On the other hand, no significant accuracy changes are observed for the proposed block-based method between {720p} and {1080p} videos.}
% When the frame-based approach is used, flat vs. natural matching scenarios always give higher TP (for FP $ =10^{-3}$) than natural vs. natural matching scenarios. This means that, in terms of scene content, flat-content videos are the most appropriate for fingerprint estimation. Moving object edges in natural-content videos imply high frequency contents that interfere with the PRNU noise estimation.}
%\textcolor{\revisionColor}{
%Therefore, on average, the number of blocks in P and B frames in which the PRNU noise is valid is bigger than the number of blocks in which the PRNU noise is invalid.
%According to previous studies like \citep{Luis2016} (which uses frame-based approach), attribution accuracy increases as the video resolution increases. Oppositely, Figures 4 to 7 show that,  when the frame-based approach is used, for a fixed FP rate ($\textrm{FP} = 10^{-3}$), TP rates obtained for  \textit{720p} videos are always higher than TP rates of  \textit{1080p} videos. 
%which uses quantization matrices to quantize coarsely high frequency DCT coefficients
%(whereas in the Main profile, the same quantization coefficient is used to quantize all the DCT coefficients of a given block) yielding a higher degradation of the PRNU noise in the encoded blocks.}

The main advantage of the proposed method is that it uses only frame blocks that have correct PRNU components during PRNU fingerprint estimation. As a result, we obtain a better PRNU fingerprint than the one estimated using the frame-based methods especially when videos are highly compressed (Table 8 and Figures 6,7). \textcolor{\revisionColor}{The results presented in this paper represent the performance lower bounds of the frame-based and block-based methods for VISION database; better source device attribution could be obtained with videos that have longer duration, specially with the block-based approach. Our future studies will focus on the evaluation of the proposed method with various video resolutions, codecs, and social media applications.}
%improving the PRNU noise estimated using block-based approach by reducing the effects of the image content (high frequency content) on the estimated PRNU noise.}
%The future research will concern the usage of this promising approach on highly compressed and digitally stabilized videos.
\section*{References}
\bibliography{mybibfile}

\end{document}